\documentclass[11pt, twocolumn]{IEEEtran}
\usepackage{enumerate}
\usepackage{ifthen}
\usepackage{amsmath,amssymb}
\usepackage{graphicx}
\usepackage{verbatim}
\usepackage{subfig}
\usepackage{ifthen}
\usepackage{multirow}
\usepackage{multicol}

\usepackage{cite}
\usepackage[usenames,dvipsnames]{color}

\newboolean{showcomments}
\setboolean{showcomments}{true}
\newcommand{\masoud}[1]{  \ifthenelse{\boolean{showcomments}}
{ \textcolor{red}{(Masoud says:  #1)}} {}  }
\newcommand{\chris}[1]{\ifthenelse{\boolean{showcomments}}
{ \textcolor{red}{(Chris says: #1)} } {} }
\newcommand{\slow}[1]{\ifthenelse{\boolean{showcomments}}
{ \textcolor{red}{(Steven says:  #1)}}{}}
\newcommand{\mani}[1]{\ifthenelse{\boolean{showcomments}}
{ \textcolor{red}{(Mani says:  #1)}}{}}

\usepackage{graphicx,amssymb,amstext,amsmath}

\usepackage{xcolor}
\usepackage{multirow}
\usepackage{lscape}
\usepackage{mdwlist}
\usepackage{url}
\usepackage{dcolumn}
\usepackage{amsfonts}
\usepackage{mathrsfs}
\usepackage{latexsym}
\usepackage{graphicx}

\usepackage[latin1]{inputenc}
\usepackage{tikz}
\usetikzlibrary{shapes,arrows}

\tikzstyle{decision} = [diamond, draw,
    text width=4.5em, text badly centered, node distance=3cm, inner sep=0pt]
\tikzstyle{block} = [rectangle, draw,
    text width=5em, text centered, rounded corners, minimum height=4em]
\tikzstyle{line} = [draw, -latex']
\tikzstyle{cloud} = [draw, ellipse, node distance=3cm,
    minimum height=2em]

\newtheorem{theorem}{Theorem}

\newtheorem{lemma}[theorem]{Lemma}
\newtheorem{corollary}[theorem]{Corollary}

\newtheorem{remark}{Remark}

\def\ba{\begin{array}}
\def\ea{\end{array}}
\newcommand{\beq}{\begin{equation}}
\newcommand{\eeq}{\end{equation}}
\newcommand{\bq}{\begin{eqnarray}}
\newcommand{\eq}{\end{eqnarray}}
\newcommand{\bqn}{\begin{eqnarray*}}
\newcommand{\eqn}{\end{eqnarray*}}
\newcommand{\bee}{\begin{enumerate}}
\newcommand{\eee}{\end{enumerate}}
\newcommand{\bi}{\begin{itemize}}
\newcommand{\ei}{\end{itemize}}

\newcommand{\btab}{\begin{tabular}}
\newcommand{\etab}{\end{tabular}}
\newcommand{\ii}{\textbf{i}}

\begin{document}

\title{Branch Flow Model: Relaxations and Convexification {\Large (Part I)}
\thanks{\textbf{To appear in {\em IEEE Trans. Power Systems}, 2013}
(submitted in May 11, 2012, accepted for publication on March 3, 2013).
A preliminary and abridged version has appeared in \cite{Farivar-2012-BFM-CDC}.}
}
\author{
Masoud Farivar \quad \quad \quad
Steven H. Low \\
Engineering and Applied Science  \\
Caltech
}
\maketitle

\begin{abstract}
We propose a branch flow model for the analysis and optimization of mesh
as well as radial networks.   The model leads to a new approach to solving optimal
power flow (OPF) that consists of two relaxation steps.   The first step
eliminates the voltage and current angles and the second step approximates the
resulting problem by a conic program that can be solved efficiently.
For radial networks, we prove that both relaxation steps are always exact, provided
there are no upper bounds on loads.   For mesh networks, the conic relaxation is
always exact but the angle relaxation may not be exact,
and we provide a simple way to determine if a relaxed solution is globally optimal.
We propose convexification of mesh networks using phase shifters so
that  OPF for the convexified network can
always be solved efficiently for an optimal solution.   We prove that convexification
requires phase shifters only outside a spanning tree of the network
and their placement depends only on network topology, not on power flows,
generation, loads, or operating constraints.
Part I  introduces our branch flow model, explains the two relaxation steps,
and proves the conditions for exact relaxation.
Part II describes convexification of mesh networks, and presents simulation
results.
%  on phase shifter ranges required for the covexification of various IEEE and other test networks.
\end{abstract}
%

% \newpage
% \tableofcontents
% \newpage

%%-------------------------------------------------------------------
% \input{intro-1.tex}
% \input{BFModel-v3.tex}
% \input{SolStrategy.tex}
% \input{ConicRelaxation.tex}
% \input{AngleRelaxation.tex}
% \input{conc-1.tex}
%%-------------------------------------------------------------------

\section{Introduction}

\subsection{Motivation}

The bus injection model is the standard model for power flow analysis and optimization.
It focuses on nodal variables such as  voltages, current and power injections and does not
directly deal with power flows on individual branches.  
Instead of nodal variables, the branch flow model focuses on currents and powers on the branches.
It has been used mainly for modeling distribution circuits
which tend to be radial, but has received far less attention.  In this paper, we advocate the use of
branch flow model for {\em both} radial and mesh networks, and demonstrate how it can be used for optimizing
the design and operation of power systems.

One of the motivations for our work is the optimal power flow (OPF) problem.
OPF seeks to optimize a certain objective function, such as power loss, generation cost and/or user utilities,
subject to Kirchhoff's laws, power balance as well as capacity, stability and contingency
constraints on the voltages and power flows.  There has been a great deal of research on OPF
since Carpentier's first formulation in 1962 \cite{Carpentier62}; surveys can be found in, e.g.,
\cite{Powerbk, Huneault91,Momoh99a,Momoh99b,Pandya08}.
OPF is generally nonconvex and NP-hard, and a large
number of optimization algorithms and relaxations have been proposed.
A popular approximation
is the DC power flow problem, which is a linearization and therefore easy to solve,
e.g. \cite{Stott1974,Alsac1990,Purchala2005,Stott2009}.
%%%
An important observation was made in \cite{bai2008, bai2009} that the full AC OPF can be
formulated as a quadratically constrained quadratic program and therefore can be
approximated by a semidefinite program.  While this approach is illustrated in \cite{bai2008,bai2009}
on several IEEE test systems using an interior-point method,
whether or when the semidefinite relaxation will turn out to be exact is not studied.
Instead of solving the OPF problem directly, \cite{Lavaei2012} proposes to solve its convex Lagrangian 
dual problem  % since the duality gap is zero when the relaxation is exact.  
and gives a sufficient condition that must be satisfied by a dual solution % for the duality gap to be zero and 
for an optimal OPF solution to be recoverable.
% Importantly, this provides a way to determine for sure if a candidate solution is globally optimal
% for the nonconvex OPF problem.    
This result is extended in \cite{Lavaei2011} to include
other variables and constraints and in \cite{Sojoudi2012PES} to exploit network sparsity.
In \cite{Bose2011,Zhang2011geometry}, it is proved that the sufficient condition of \cite{Lavaei2012}
always holds for a radial (tree) network, provided the bounds on the power flows satisfy a simple
pattern.  See also \cite{Bose-2012-QCQPt} for a generalization.  
These results confirm that radial networks are computationally much simpler.
This is important as most distribution systems are radial.

The limitation of semidefinite relaxation for OPF  is
studied in \cite{Lesieutre-2011-OPFSDP-Allerton}  using mesh networks with
3, 5, and 7 buses: 
as a line-flow constraint is tightened, % the sufficient condition in \cite{Lavaei2012}
% fails to hold for these examples and 
the duality gap becomes nonzero and the solutions
produced by the semidefinite relaxation becomes physically meaningless.  % in those cases. 
Indeed, examples of nonconvexity have long been discussed in the literature, e.g.,
\cite{Hiskens-2001-OPFboundary-TPS, LesieutreHiskens-2005-OPF-TPS, Hill-2008-OPFboundary-TPS}.
% Hence it is important to develop systematic methods for solving OPF involving mesh networks when 
% convex relaxation fails.
See, e.g., \cite{Phan2012} for branch-and-bound algorithms for solving OPF when convex relaxation
fails.

The papers above are all based on the bus injection model.  In this paper, we introduce a branch flow
model on which OPF and its relaxations can also be defined.   Our model is motivated by a model
first proposed by Baran and Wu in \cite{Baran1989a, Baran1989b} for the optimal placement and sizing
of switched capacitors in distribution circuits for Volt/VAR control.
One of the  insights we highlight here is that the Baran-Wu model of
 \cite{Baran1989a, Baran1989b} can be treated as a particular relaxation of our branch flow model where
 the phase angles of the voltages and currents are ignored.
By recasting their model as a set of linear and quadratic equality constraints, 
\cite{Farivar2011-VAR-SGC, Farivar2012-VAR-PES} observe that relaxing the quadratic equality 
constraints to inequality constraints yields a second-order cone program (SOCP).  
It proves that the SOCP relaxation  is exact for radial networks, when there are no
upper bounds on the loads.   This result is extended here to mesh networks with line
limits, and convex, as opposed to linear, objective functions (Theorem \ref{thm:ecr}).
See also \cite{Taylor2011PhD, Taylor2012-TPS} for various convex relaxations of approximations
of the Baran-Wu model for radial networks.

Other branch flow models have also been studied, e.g.,  in \cite{Cespedes1990, Exposito1999, Jabr2006},
all for radial networks.
Indeed \cite{Cespedes1990} studies a similar model to that in \cite{Baran1989a, Baran1989b},
using receiving-end branch powers as variables instead of sending-end branch powers
as in \cite{Baran1989a, Baran1989b}.
Both \cite{Exposito1999} and  \cite{Jabr2006} eliminate voltage angles
by defining real and imaginary parts of $V_iV_j^*$ as new variables and defining 
bus power injections in terms of these new variables.  This results in a system of
linear quadratic equations in power injections and the new variables.   
While \cite{Exposito1999} develops a Newton-Raphson algorithm to solve the bus power injections,
\cite{Jabr2006}  solves for the branch flows through an SOCP relaxation for 
radial networks, though no proof of optimality is provided.

This set of papers 
\cite{Baran1989a, Baran1989b, Cespedes1990, Exposito1999, Jabr2006, Taylor2011PhD, Farivar2011-VAR-SGC,
Taylor2012-TPS, Farivar2012-VAR-PES}
all exploit the fact that power flows can be specified by a simple set 
of linear and quadratic equalities if voltage angles can be eliminated.     Phase angles can
be relaxed only for radial networks and generally not for mesh networks, 
as \cite{Jabr2007} points out for their branch flow model, because cycles in a mesh network
impose nonconvex constraints on the optimization variables (similar to the angle
recovery condition in our model; see Theorem \ref{thm.2} below).    
For mesh networks, \cite{Jabr2007} proposes a sequence of  
SOCP where the nononvex constraints are replaced by their linear approximations and demonstrates 
the effectiveness of this approach using seven network examples.   
In this paper we extend the Baran-Wu model from radial to mesh networks and use it to develop a 
solution strategy for OPF.

\subsection{Summary}

Our purpose is to develop a formal theory of branch flow model for the analysis and optimization of
mesh as well as radial networks.   As an illustration, we formulate OPF within this alternative model, 
propose relaxations, characterize when a relaxed solution is exact, prove that our relaxations are 
always exact for radial networks when there are no upper bounds on loads but may not be exact
for mesh networks, and show how to use phase shifters to convexify a mesh network so that a 
relaxed solution is always optimal for the convexified network.   
% A similar set of results have been proved in the 
% sequence of papers \cite{Lavaei2012, Bose2011, Zhang2011geometry, Sojoudi2012PES, Bose-2012-QCQPt} for the 
% bus injection model, even though the results have distinct characters in each model and the 
% proof techniques are completely different. 
% Indeed, it can be shown that the two models are equivalent  and both help deepen 
% our understanding of OPF. 

Specifically we  formulate in Section \ref{sec:bfm} the OPF problem using branch flow equations
involving complex bus voltages and complex branch current and power flows.
%%%%%%%%%%%%%%%%
\begin{figure}[htbp]
\centering
\includegraphics[width=0.5\textwidth]{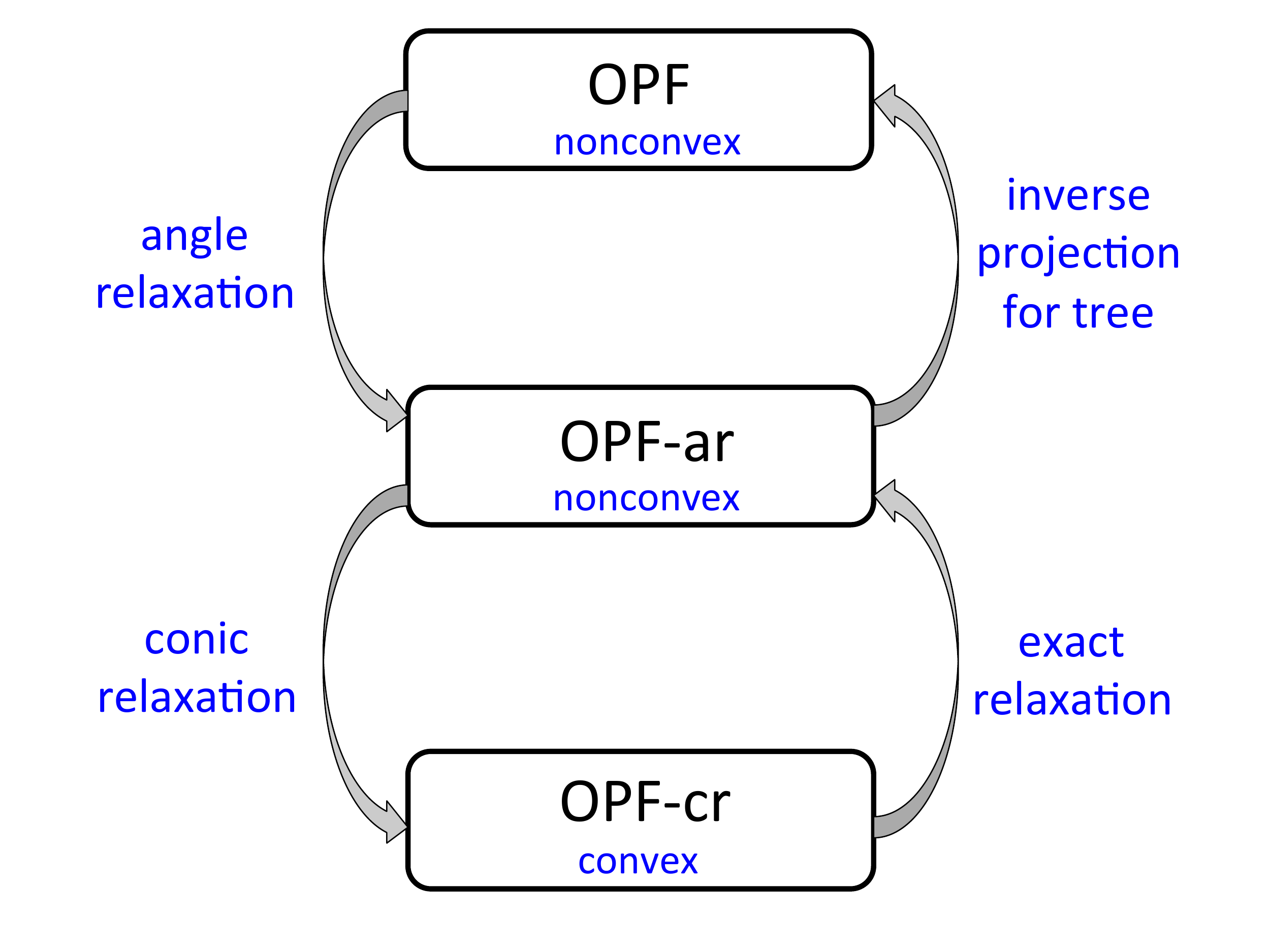}
\caption{Proposed solution strategy for solving OPF.  
}
\label{fig:strategy1}
\end{figure}
%%%%%%%%%%%%%%%%
In Section \ref{sec:rss} we describe our solution
approach that  consists of two relaxation steps (see Figure \ref{fig:strategy1}):
\bi
\item {\em Angle relaxation}: relax OPF by eliminating voltage and current angles from the branch flow
	equations.  This yields the (extended) Baran-Wu model and a relaxed problem OPF-ar which is still 
	nonconvex due to a quadratic equality constraint.
\item {\em Conic relaxation}: relax OPF-ar by changing the quadratic equality into an inequality constraint.  
	This yields a convex problem OPF-cr (which is an SOCP when the objective function is linear).
%	 that can be solved efficiently.
\ei

\begin{figure}[htbp]
\centering
\includegraphics[width=0.4\textwidth]{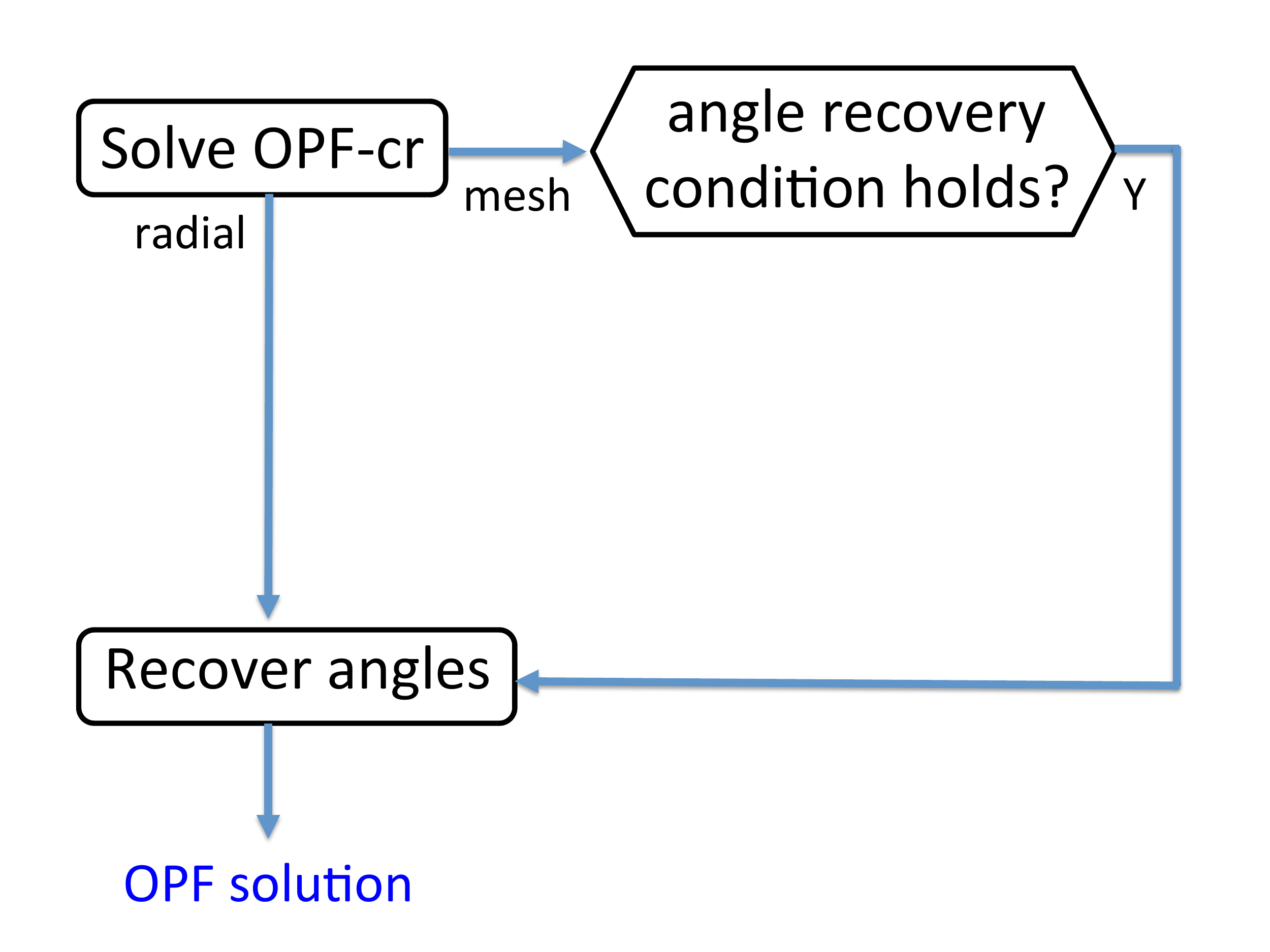}
\caption{Proposed algorithm for solving OPF (\ref{eq:opf1.a})--(\ref{eq:opf1.c})
without phase shifters.  
The details are explained in Sections \ref{sec:bfm}--\ref{sec:ar}.}
\label{fig:strategy}
\end{figure}

In Section \ref{sec:cr} we prove that the conic relaxation OPF-cr is always exact  {\em even for}
mesh networks, provided there are no upper bounds on real and reactive loads, i.e., 
{\em any} optimal solution of OPF-cr is also optimal for OPF-ar.  
Given an optimal solution of OPF-ar, whether we can derive  an optimal solution of the original OPF
depends on whether we can recover the voltage and current angles  from the given OPF-ar
solution.  
In Section \ref{sec:ar} we characterize the exact condition (the angle recovery condition)
under which this is possible, and present two angle recovery algorithms.
The angle recovery condition has a simple interpretation: any  solution
of OPF-ar implies an angle difference across a line, and the condition 
says that the implied  angle differences sum to zero (mod $2\pi$) around each cycle.
For a radial network, this condition holds trivially and hence 
solving the conic relaxation OPF-cr always produces an optimal solution for OPF.
For a mesh network, the angle recovery condition corresponds to the requirement that the implied
phase angle differences sum to zero around every loop.   The given OPF-ar solution may not
satisfy this condition, but our characterization can be used to check if it yields an optimal solution for OPF.
These results suggest an algorithm for solving OPF as summarized in Figure \ref{fig:strategy}.
%%%%%%%%%%%%%%%%
%%%%%%%%%%%%%%%%

If a relaxed solution for a mesh network does not satisfy the angle recovery condition,
then it is  infeasible  for OPF.
In Part II of this paper, we propose a simple way to convexify a mesh network using phase
shifters so that {\em any} relaxed solution of OPF-ar can be mapped to an optimal solution of 
OPF for the convexified network, with an optimal cost that is lower than or equal to that of the 
original network.

% \slow{Discuss somewhere:
% The solution of power flow (also called load
% flow) equations are studied in e.g., \cite{TavoraSmith1972,SastryVaraiya1981,Baillieul1982,ArapostathisVaraiya1983,Thorp1986}.
% While many numerical solution methods have been developed, scant attention has been
% devoted to understanding structural properties of power flow equations, such as the existence
% and uniqueness of solution, stability of solutions (their Jacobian matrix being positive definite),
% and the geometry of the solution sets.  }

\vspace{-0.1in}
\subsection{Extensions: radial networks and equivalence}

In \cite{Gan-2012-BFMt, Li-2012-BFMt}, we prove a variety of sufficient conditions under which the conic
 relaxation proposed here is exact for radial networks.  The main difference from Theorem~\ref{thm:ecr} below
 is that, \cite{Gan-2012-BFMt, Li-2012-BFMt} allow upper bounds on the loads but relax upper bounds on
 voltage magnitudes.   Unlike the proof for Theorem \ref{thm:ecr} here, those in 
 \cite{Gan-2012-BFMt, Li-2012-BFMt} exploit the duality theory.
  
The bus injection model and the branch flow model are defined by different sets of equations in terms of 
their own variables.   Each model is self-contained: one can formulate and analyze power flow problems 
within each model, using only nodal variables or only branch variables.
Both models (i.e., the sets of equations in their respective variables), however, are descriptions of the Kirchhoff's
laws. 
In \cite{Bose-2012-BFMe-Allerton} we prove formally the equivalence of these models, in the sense
that given a power flow solution in one model, one can derive a corresponding power flow solution
in the other model.
Although the semidefinite relaxation in the bus injection model is very different from the convex relaxation
proposed here, \cite{Bose-2012-BFMe-Allerton} also establishes the precise relationship between the 
various relaxations in these two models.
%
% As a consequence, the various conditions on radial 
% networks proved in \cite{Bose2011, Zhang2011geometry, Sojoudi2012PES, Bose-2012-QCQPt} for 
% the exactness of semidefinite relaxation 
% imply immediately the exactness of the relaxations proposed here.  Conversely, 
% the conditions in \cite{Gan-2012-BFMt, Li-2012-BFMt} for the conic relaxation immediately imply
% that there is a matrix solution of the semidefinite relaxation in 
% \cite{Bose2011, Zhang2011geometry, Sojoudi2012PES, Bose-2012-QCQPt} that is of rank 1.  
This is useful because some results are easier to formulate and prove in one model than
in the other.   For instance, it is hard to see how the upper bounds on voltage magnitudes and
the technical conditions on the line impedances in \cite{Gan-2012-BFMt, Li-2012-BFMt} for exactness 
in the branch flow model affect the rank of the semidefinite matrix variable in the bus injection model,
although \cite{Bose-2012-BFMe-Allerton} clarifies conditions that guarantee their equivalence.

% e.g., BFM deals with physical quantities directly (branch flows, branch currents, 
% voltages) that often have intuitive interpretation, whereas, for SDP relaxation in BIM, the variables 
% are semideifinite matrices that are less intuitive.

\section{Branch flow model}
\label{sec:bfm}

Let $\mathbb{R}$ denote the set of real numbers, $\mathbb{C}$ complex numbers,
and $\mathbb{N}$  integers.
A variable without a subscript  denotes a vector with appropriate components,
e.g., $s := (s_i, i=1, \dots, n)$, $S := (S_{ij}, (i,j)\in E)$.   
% For a complex scalar or vector $a$, $a^*$ denotes its complex conjugate.
For a vector $a = (a_1, \dots, a_k)$, $a_{-i}$ denotes $(a_1, \dots, a_{i-1}, a_{i+1}, a_k)$.
For a scalar, vector, or matrix $A$, $A^t$ denotes its transpose and $A^*$ its complex conjugate transpose.
Given a directed graph $G = (N, E)$,   
denote a link in $E$ by $(i, j)$ or $i\rightarrow j$ if it points from node $i$ to
node $j$.    We will use $e$, $(i, j)$, or $i\rightarrow j$ interchangeably to refer to a link in $E$.
We write $i\sim j$ if $i$ and $j$ are connected, i.e., if either $(i, j)\in E$ or $(j, i)\in E$ (but not both).
% A  {\em cycle} $c$ in $G$ is an {\em ordered} list $c = (i_1, \dots, i_k)$ of nodes in $N$ such that 
% $i_1\sim i_2, \dots, i_k \sim i_1$.  These are links in $E$ but each link may be in the same
% orientation (as defined by the order of the nodes in $c$) or in the opposite orientation.
% We will use `$(i,j)\in c$' to denote a link $i\sim j$ in the cycle $c$.
% and `$i\in c$' to denote a node $i$ in $c$.
We write $\theta = 0$ (mod $2\pi$) if $\theta = 2\pi k$,
and $\theta = \phi$ (mod $2\pi$) if $\theta - \phi = 2\pi k$, for some integer $k$.
 For an $d$-dimensional vector $\alpha$,  $\mathcal{P}(\alpha)$ denotes its projection
 onto $(-\pi, \pi]^d$ by taking modulo $2\pi$ componentwise.

%%%%%%%%%%%%%%%%%%%%%%%%%%%%%%%
\subsection{Branch flow model}
\label{sec:branchflow}

Let $G = (N, E)$  be a connected graph representing a power network, where each
node in $N$\ represents a bus and each link in $E$ represents a line (condition A1).
We index the nodes by $i = 0, 1, \dots, n$.
The power network is called {\em radial} if its graph $G$ is a tree.
For a distribution network, which is typically radial, the root  of the tree (node 0)
represents the substation bus.
For a (generally meshed) transmission network, node 0 represents the slack bus.
% We use node $n$ to represent ground so that if bus $i$ has a shunt impedance,
% then node $i$ is connected to node $n$, i.e., $(i, n)\in E$.

We regard $G$ as a directed graph and adopt the following orientation for convenience (only).
Pick {\em any} spanning tree $T := (N, E_T)$ of $G$  rooted at node 0,
i.e., $T$ is connected and $E_T \subseteq E$ has $n$ links.  
All links in $E_T$ point away from the root.   For any link in $E\setminus E_T$ that is not
in the spanning tree $T$, pick an arbitrary direction.    
Denote a link by $(i, j)$ or $i\rightarrow j$ if it points from node $i$ to
node $j$.    
% We will use $e$, $(i, j)$, or $i\rightarrow j$ interchangeably to refer to a link in $E$.
% We write $i\sim j$ if $i$ and $j$ are connected, i.e., if either $(i, j)\in E$ or $(j, i)\in E$ (but not both).
% For each link $(i, j)\in E$, we will call node $i$ the {\em parent} of node $j$ and $j$ the {\em child} of $i$.
% Let $\pi(j) \subseteq N$ be the set of all parents of node $j$ and $\delta(i) \subseteq N$ the set
% of all children of node $i$.
Henceforth we will assume without loss of generality that $G$ and $T$ are directed graphs
as described above.\footnote{The orientation of $G$ and $T$ are different for different spanning
trees $T$, but we often ignore this subtlety in this paper.}
For each link $(i,j)\in E$, 
let $z_{ij} = r_{ij} + \ii x_{ij}$ be the complex impedance on the line,
and $y_{ij} := 1/z_{ij} =: g_{ij} - \ii b_{ij}$ be the corresponding admittance.
For each node $i \in N$, let $z_{i} = r_{i} + \ii x_i$ be the shunt impedance from $i$ to ground,
and $y_i := 1/z_i =: g_i - \ii b_i$.\footnote{The shunt admittance $y_i$ represents
 capacitive devices on bus $i$ only and a line is modeled by a series admittance $y_{ij}$ without
 shunt elements.  If a shunt admittance $\ii \tilde{b}_{ij}/2$ is included on each end of  line
 $(i,j)$ in the $\pi$-model, then a limit on line flow should be a limit on 
 $\left|S_{ij} - \ii \tilde{b}_{ij}|V_i|^2/2\right|$ instead of on $|S_{ij}|$.}

For each $(i, j)\in  E$,
let $I_{ij}$ be the complex current from buses $i$ to $j$ and 
$S_{ij} = P_{ij} + \ii Q_{ij}$ be the {\em sending-end} complex power from buses $i$ to $j$.
For each node $i\in N$, let $V_i$ be the complex voltage on bus $i$.
Let $s_i$ be the net complex power injection, which is generation minus load on bus $i$.
We use $s_i$ to denote both the complex number $p_i + \ii q_i$ and the pair $(p_i, q_i)$
depending on  the context.

As customary, we assume that the complex voltage $V_0$ is
given and the complex net generation $s_0$ is a variable.
For power flow analysis, we assume other power injections $s := (s_i, i= 1, \dots, n)$ are given.  
For optimal power flow, VAR control, or demand response, $s$ are control 
variables as well.

Given $z := (z_{ij}, (i,j)\in E, \ z_i, i\in N)$, $V_0$ and bus power injections $s$,
the variables $(S, I, V, s_0) := (S_{ij}, I_{ij}, (i, j)\in  E, \ V_i, i = 1, \dots, n, \ s_0)$ satisfy
the Ohm's law:
\bq
V_i - V_j & = & z_{ij}I_{ij},
\ \ \ \ \ \ \forall (i, j)\in E
\label{eq:Kirchhoff.1b}
\eq
the definition of branch power flow:
\bq
S_{ij} & = & V_i I_{ij}^*,
\ \ \ \ \ \ \forall (i, j)\in E
\label{eq:Kirchhoff.1c}
\eq
and power balance at each bus: for all $j\in N$,
\bq
\!\!\!\!\!\!
\sum_{k: j \rightarrow k} S_{jk} - \!\! \sum_{i: i\rightarrow j} \!\! \left( S_{ij} - z_{ij} |I_{ij}|^2 \right) + y_j^* |V_j|^2 
& \!\!\! = \!\!\! & s_j
\label{eq:Kirchhoff.1a}
\eq
We will refer to  \eqref{eq:Kirchhoff.1b}--\eqref{eq:Kirchhoff.1a} as the
{\em branch flow model/equations}.
Recall that the cardinality $|N| = n+1$ and let $ | E |=:m$.
The branch flow equations (\ref{eq:Kirchhoff.1b})--(\ref{eq:Kirchhoff.1a}) specify $2m+n+1$
nonlinear equations in $2m+n+1$
complex variables $(S, I, V, s_0)$, when other bus power injections $s$ are specified.

We will call a solution of (\ref{eq:Kirchhoff.1b})--(\ref{eq:Kirchhoff.1a}) a {\em branch flow solution}
with respect to a given $s$, and denote it by $x(s) := (S, I, V, s_0)$.
Let $\mathbb{X}(s) \subseteq \mathbb{C}^{2m+n+1}$ be the set of all branch flow solutions
with respect to a given $s$:
\bq
\mathbb X(s) & \!\! \!\! := &\!\!\!\! \left\{ x := (S, I, V, s_0) \, |\, x \text{ solves 
		(\ref{eq:Kirchhoff.1b})--(\ref{eq:Kirchhoff.1a}) given $s$} \right\}
\nonumber
\\
\label{eq:defXs}
\eq
and let $\mathbb X$ be the set of all branch flow solutions:
\bq
\mathbb X & := & \bigcup_{s\in \mathbb C^n}\  \mathbb X(s)
\label{eq:defX}
\eq
For simplicity of exposition, we will often abuse notation and use $\mathbb X$ to denote either the
set defined in \eqref{eq:defXs} or that in \eqref{eq:defX}, depending on the context.  For
instance, $\mathbb X$ is used to denote the set in \eqref{eq:defXs} for a fixed $s$ in Section
\ref{sec:ar} for power flow analysis, and to denote the set in \eqref{eq:defX} in Section
\ref{sec:cr} for optimal power flow where $s$ itself is also an optimization variable.
Similarly for other variables such as $x$ for $x(s)$.

%%%%%%%%%%%%%%%%%%%%%%%%%%%%%%%
\subsection{Optimal power flow}
\label{sec:opf}

Consider the optimal power flow problem where, in addition to $(S, I, V, s_0)$,
 $s$ is also an optimization variable.
Let $p_i := p_i^g - p_i^c$ and $q_i := q_i^g - q_i^c$ where
$p_i^g$ and $q_i^g$ ($p_i^c$ and $q_i^c$) are the real and reactive power generation
(consumption) at node $i$.
For instance, \cite{Baran1989a,Baran1989b} formulate a Volt/VAR control problem for a distribution circuit
where $q_i^g$ represent the placement and sizing of shunt capacitors.
In addition to  (\ref{eq:Kirchhoff.1b})--(\ref{eq:Kirchhoff.1a}), we impose the following constraints
on power generation: for $i \in N$,
\begin{equation}
\underline{p}_i^g \leq p_i^g \leq \overline{p}_i^g, \quad \underline{q}_i^g \leq q_i^g  \leq \overline{q}_i^g
\label{GenLimits}
\end{equation}
In particular, any of $p_i^g, q_i^g$ can be a fixed constant by specifying that
$\underline{p}_i^g = \overline{p}_i^g$ and/or $\underline{q}_i^g = \overline{q}_i^g$.
For instance, in the inverter-based VAR control problem of \cite{Farivar2011-VAR-SGC, Farivar2012-VAR-PES}, 
$p_i^g$ are the fixed (solar)
power outputs and the reactive power $q_i^g$ are the control variables.
For power consumption, we require, for $i\in N$,
\begin{equation}
\underline{p}_i^c \leq p_i^c \leq \overline{p}_i^c, \quad \underline{q}_i^c \leq q_i^c \leq \overline{q}_i^c
\label{oversatisfaction}
%\label{OPF1}
\end{equation}
The voltage magnitudes must be maintained in tight ranges: for $i=1, \dots, n$,
%for all $i\not\in V^\prime$,
\begin{eqnarray}
\underline{v}_i & \leq \ \ |V_i|^2 \ \ \leq  & \overline{v}_i
 \label{Vlimits}
 \end{eqnarray}
 Finally, we impose  flow limits in terms of branch currents: for all $(i,j)\in E$,
 \bq
 |I_{ij}| & \leq & \overline{I}_{ij}
 \label{eq:ll}
 \eq

We allow any objective function that is convex and does not depend on the angles 
$\angle V_i, \angle I_{ij}$ of voltages and currents.
For instance, suppose we aim to minimize real power losses $r_{ij}  |I_{ij}|^2$ \cite{Nazari2006, Nazari2010}, minimize real power generation costs $c_i p_i^g$,   and maximize
energy savings through conservation voltage reduction (CVR).  
 Then the objective function takes the form (see \cite{Farivar2011-VAR-SGC, Farivar2012-VAR-PES})
\bq
\sum_{(i,j)\in E} r_{ij}  |I_{ij}|^2 + \sum_{i\in N} c_i p_i^g + \sum_{i\in N}{\alpha_i  |V_i|^2}
\label{eq:egf}
\eq
for some given constants $c_i, \alpha_i\geq 0$.
% We also allow the cost to be quadratic in real power.

To simplify notation, let $\ell_{ij}:= |I_{ij}|^2$ and $v_i := |V_i|^2$.
Let $s^g := (s_i^g, \, i=1, \dots, n) = (p_i^g, q_i^g, \, i=1, \dots, n)$ be the power generations,
and $s^c := (s_i^c, \, i=1, \dots, n) = (p_i^c, q_i^c, \, i=1, \dots, n)$ the power consumptions.
Let  $s$ denote either $s^g - s^c$ or $(s^g, s^c)$ depending on the context.
Given a branch flow solution $x := x(s) := (S, I, V, s_0)$ with respect to a given $s$, let 
$\hat{y} := \hat{y}(s) := (S, \ell, v, s_0)$ denote the projection of $x$ that have phase angles 
$\angle V_i, \angle I_{ij}$ eliminated.   This defines a projection function $\hat{h}$ such that
 $\hat{y} = \hat{h}(x)$,
to which we will return in Section \ref{sec:rss}.   Then our objective function is 
$f\left(\hat{h}(x), s \right)$.
We assume $f\left( \hat{y}, s \right)$ is convex
(condition A2); in addition, we assume $f$ is strictly increasing in $\ell_{ij}, (i, j)\in E$,
 nonincreasing in load $s^c$, and independent of $S$ (condition A3).   Let
 \bqn
 \mathbb S & := & \{\, (S, v, s_0, s) \, |\, (v, s_0, s) \text{ satisfies } (\ref{GenLimits})-(\ref{eq:ll})\, \}
 \eqn
All quantities are optimization variables, except $V_0$ which is given.

The optimal power flow problem is
\\
\noindent
\textbf{OPF}:
\bq
\min_{x, s} & & f \left( \hat{h}(x), s \right)
\label{eq:opf1.a}
\\
\text{subject to} & &  x \in \mathbb X, \quad (S, v, s_0, s) \in \mathbb S
\label{eq:opf1.c}
\eq
where $\mathbb X$ is defined in (\ref{eq:defX}).
% To avoid triviality, we assume the problem is feasible (condition A4).

The feasible set is specified by the nonlinear branch flow equations
 and hence OPF (\ref{eq:opf1.a})--(\ref{eq:opf1.c})
 is in general nonconvex and hard to solve.
The goal of this paper is to propose an efficient way to solve OPF
by exploiting the structure of the branch flow model.

%%%%%%%%%%%%%%%%%%%%%%%%%%%%%%%

\subsection{Notations and assumptions}

The main variables and assumptions are summarized in Table \ref{table1:notations} and below
for ease of reference:
\begin{table}[htbp]
\caption{Notations.}
\centering
\begin{tabular}{|| l | p{5cm} ||}
\hline \hline
	$G$, $T$ & (directed) network graph $G$ and a spanning tree $T$ of $G$ \\
\hline
	$B$, $B_T$ & reduced (and transposed) incidence matrix of $G$ and the submatrix corresponding to $T$	\\
\hline
	$V_i$, $v_i$ & complex voltage on bus $i$ with $v_i :=  |V_i|^2$\\
\hline
	$s_i = p_i + \ii q_i$ & net complex load power on bus $i$ \\
	$p_i = p_i^g - p_i^c$ & net real power equals generation minus load; \\
	$q_i = q_i^g - q_i^c$ & net reactive power equals generation minus load \\
\hline
	$I_{ij}$, $\ell_{ij}$ & complex current from buses $i$ to $j$ with $\ell_{ij}:= |I_{ij}|^2$  \\
\hline

	$S_{ij} = P_{ij} + \ii Q_{ij}$ & complex power from buses $i$ to $j$  (sending-end)  \\
\hline
	$\mathbb X$  &  set of all branch flow solutions that satisfy 
		\eqref{eq:Kirchhoff.1b}--\eqref{eq:Kirchhoff.1a} either for some $s$, or for a given $s$
		(sometimes denoted more accurately by $\mathbb X(s) )$;  \\
	$\hat{\mathbb Y}$  & set of all relaxed branch flow solutions that satisfy 
			\eqref{eq:Kirchhoff.2a}--\eqref{eq:Kirchhoff.2d} either for a given $s$ or for some $s$;   \\
	$\overline{\mathbb Y}$ & set of all relaxed branch flow solutions that satisfy 
	 		(\ref{eq:Kirchhoff.2a})--(\ref{eq:Kirchhoff.2c}) and \eqref{eq:opf3.d}
			either for a given $s$ or for some $s$;  \\

%	$\overline{X}$, $\overline{X}_T$  &  set of branch flow solutions that satisfy \eqref{eq:Kirchhoff.1a}, 
%			\eqref{eq:Kirchhoff.3b}, \eqref{eq:Kirchhoff.1c}, for some phase shifter angles $\phi$
%			and for some $\phi \in T^\perp$;	\\
			
\hline
	$x = (S, I, V, s_0) \in \mathbb X$  &  vector $x$ of power flow variables \\
	$\hat{y} = (S, \ell, v, s_0) \in \hat{\mathbb Y}$  &  and its projection $\hat{y}$;    \\ 
	$\hat{y} = \hat{h}(x); \ \ x = h_{\theta}(\hat{y})$ &  projection mapping $\hat{y}$ and an inverse $h_\theta$ \\

\hline
	$z_{ij}$, $y_i$ & impedance on line $(i,j)$  and shunt admittance
			from bus $i$ to ground \\
\hline
	$f =  f \left( \hat{h}(x), s \right)$  &   objective function of OPF  \\
% \hline
% 	$a_{-i}$ & $(a_1, \dots, a_{i-1}, a_{i+1}, a_k)$ \\
% \hline
%	$a^*$ & complex conjugate of $a$ \\
\hline \hline
\end{tabular}
\label{table1:notations}
\end{table}

\bee
\item[A1]  The network graph $G$ is connected.
\item[A2]  The cost function $f(\hat{y}, s)$ for optimal power flow is convex.  
\item[A3]  The cost function $f(\hat{y}, s)$ is strictly increasing in $\ell$,
		 nonincreasing in load $s^c$, and independent of $S$.
\item[A4]  The optimal power flow problem OPF (\ref{eq:opf1.a})--(\ref{eq:opf1.c}) is feasible.
\eee
These assumptions are standard and realistic.
For instance, the objective function in \eqref{eq:egf} satisfies conditions A2--A3.
 A3 is a property of the objective function $f$  and not a property of  power 
 flow solutions; it holds if the cost function is strictly increasing in line loss.

%%%%%%%%%%%%%%%%%%%%%%%%%%%%%%%
\section{Relaxations and solution strategy}
\label{sec:rss}

% We now describe our solution approach.

%%%%%%%%%%%%%%%%%%%%%%%%%%%%%%%%%%%%%%%%%%%%%%%%%%%%
\subsection{Relaxed branch flow model}

Substituting (\ref{eq:Kirchhoff.1c}) into (\ref{eq:Kirchhoff.1b}) yields  $V_j = V_i - z_{ij}S_{ij}^*/V_i^*$.
Taking the magnitude squared, we have $v_j =  v_i + |z_{ij}|^2 \ell_{ij} - (z_{ij} S_{ij}^* + z_{ij}^* S_{ij})$.
Using (\ref{eq:Kirchhoff.1a}) and (\ref{eq:Kirchhoff.1c}) and in terms of real variables, we therefore
have
\bq
p_j & \!\!\! = \!\!\! & \!\!\! 
\sum_{k: j\rightarrow k} \!\! P_{jk} - \!\!\! \sum_{i:i\rightarrow j} \!\! \left( P_{ij} - r_{ij} \ell_{ij} \right) +  g_j v_j, 
\ \forall j
\label{eq:Kirchhoff.2a}
\\
q_j & \!\!\! = \!\!\! & \!\!\! 
\sum_{k: j\rightarrow k} \!\! Q_{jk} - \!\!\! \sum_{i:i\rightarrow j} \!\! \left( Q_{ij} - x_{ij} \ell_{ij} \right) +  b_j v_j, 
\ \forall j
\label{eq:Kirchhoff.2b}
\\
v_j & = & v_i - 2 (r_{ij} P_{ij} + x_{ij} Q_{ij}) + (r_{ij}^2 + x_{ij}^2) \ell_{ij}
\nonumber
\\
& & \ \ \ \ \ \ \ \ \  \ \ \ \ \ \ \ \ \ \ \ \ \ \ \ \ \
\ \forall (i, j)\in E
\label{eq:Kirchhoff.2c}
\\
\ell_{ij} & = & \frac{P_{ij}^2 + Q_{ij}^2}{v_i},
\ \ \ \ \ \ \ \ \ \ \  \ \ \ \forall (i, j)\in E
\label{eq:Kirchhoff.2d}
\eq
We will refer to (\ref{eq:Kirchhoff.2a})--(\ref{eq:Kirchhoff.2d}) as the {\em relaxed (branch flow)
model/equations} and a solution a {\em relaxed (branch flow) solution}.
These equations were first proposed in \cite{Baran1989a,Baran1989b} to model radial
distribution circuits.
They define a system of equations in the variables
$(P, Q, \ell, v, p_0, q_0) := (P_{ij}, Q_{ij}, \ell_{ij}, (i, j)\in E, \
v_{i}, i=1, \dots, n, \ p_0, q_0)$.
 We often use  $(S, \ell, v, s_0)$ as a shorthand for $(P, Q, \ell, v, p_0, q_0)$. 
The relaxed model has a solution under A4.

In contrast to the original branch flow equations (\ref{eq:Kirchhoff.1b})--(\ref{eq:Kirchhoff.1a}),
the relaxed equations (\ref{eq:Kirchhoff.2a})--(\ref{eq:Kirchhoff.2d}) specifies $2(m+n+1)$
equations in $3m + n + 2$ real variables $(P, Q, \ell, v, p_0, q_0)$,   given $s$.
For a radial network, i.e., $G$ is a tree,
$m = |E| = |N|-1 = n$.
Hence the relaxed system  (\ref{eq:Kirchhoff.2a})--(\ref{eq:Kirchhoff.2d})
specifies $4n+2$ equations in $4n+2$ real variables.
It is shown in \cite{ChiangBaran1990} that there are generally multiple solutions, but
for practical networks where $|V_0| \simeq 1$ and $r_{ij}, x_{ij}$ are small p.u., the solution
of (\ref{eq:Kirchhoff.2a})--(\ref{eq:Kirchhoff.2d}) is unique.
Exploiting structural properties of the Jacobian matrix, efficient algorithms have also been proposed in
\cite{Chiang1991} to solve the relaxed branch flow equations.

For a connected mesh network, $m=|E| > |N|-1=n$, in which case there are more
variables than equations for the relaxed model
(\ref{eq:Kirchhoff.2a})--(\ref{eq:Kirchhoff.2d}), and therefore the solution
 is generally nonunique.    Moreover, some of these solutions may be spurious, i.e., they do not
correspond to a solution of the original branch flow equations (\ref{eq:Kirchhoff.1b})--(\ref{eq:Kirchhoff.1a}).

Indeed, one may consider  $(S, \ell, v, s_0)$
as a projection of $(S, I, V, s_0)$ where each variable $I_{ij}$ or $V_i$ is relaxed from a point in the
 complex plane to a circle with a radius equal to the distance of the point from the origin.  It is therefore not 
 surprising that a relaxed solution
 of (\ref{eq:Kirchhoff.2a})--(\ref{eq:Kirchhoff.2d}) may not correspond to
 any solution of (\ref{eq:Kirchhoff.1b})--(\ref{eq:Kirchhoff.1a}).
 The key is whether, given a relaxed solution, we can recover the angles $\angle V_i, \angle I_{ij}$ 
 correctly from it.     It is then remarkable that, when $G$ is a tree,
 indeed the solutions of (\ref{eq:Kirchhoff.2a})--(\ref{eq:Kirchhoff.2d}) coincide with those of
 (\ref{eq:Kirchhoff.1b})--(\ref{eq:Kirchhoff.1a}).
Moreover {\em for a general network, (\ref{eq:Kirchhoff.2a})--(\ref{eq:Kirchhoff.2d}) together with
the angle recovery condition in Theorem \ref{thm.2} below are indeed equivalent to 
 (\ref{eq:Kirchhoff.1b})--(\ref{eq:Kirchhoff.1a})}, as explained in Remark \ref{remark:eq}
 of Section \ref{sec:ar}.

To understand the relationship between the branch flow model and the relaxed model
and formulate our relaxations precisely, we need some notations.    Fix an $s$.
 % Let $x := (S, I, V, s_0)$ denote a point in $\mathbb{C}^{2m+n+1}$ and
 % $y := (S, \ell, v, s_0) := (P, Q, \ell, v, p_0, q_0)$ a point in $\mathbb{R}^{3m+n+2}$.
  %
 Given a vector $(S, I, V, s_0) \in \mathbb{C}^{2m+n+1}$, define its projection
 $\hat{h}: \mathbb{C}^{2m+n+1} \rightarrow \mathbb{R}^{3m+n+2}$ by
 $\hat{h}(S, I, V, s_0)  =  (P, Q, \ell, v, p_0, q_0)$ where
 \bq
P_{ij} = \text{Re} \ S_{ij}, &  & Q_{ij} = \text{Im} \ S_{ij}, \ \ \ell_{ij} = |I_{ij}|^2
\label{eq:defhhat.1}
\\
% \ell_{ij} = |I_{ij}|^2, & & v_i = |V_i|^2
% \label{eq:defhhat.2}
% \\
p_i = \text{Re} \ s_i,  & & q_i  = \text{Im} \ s_i, \ \  \ \  v_i = |V_i|^2
\label{eq:defhhat.3}
\eq
%%%%%%%%%%%%%%%%
\begin{figure}[htbp]
\centering
\includegraphics[width=0.33\textwidth]{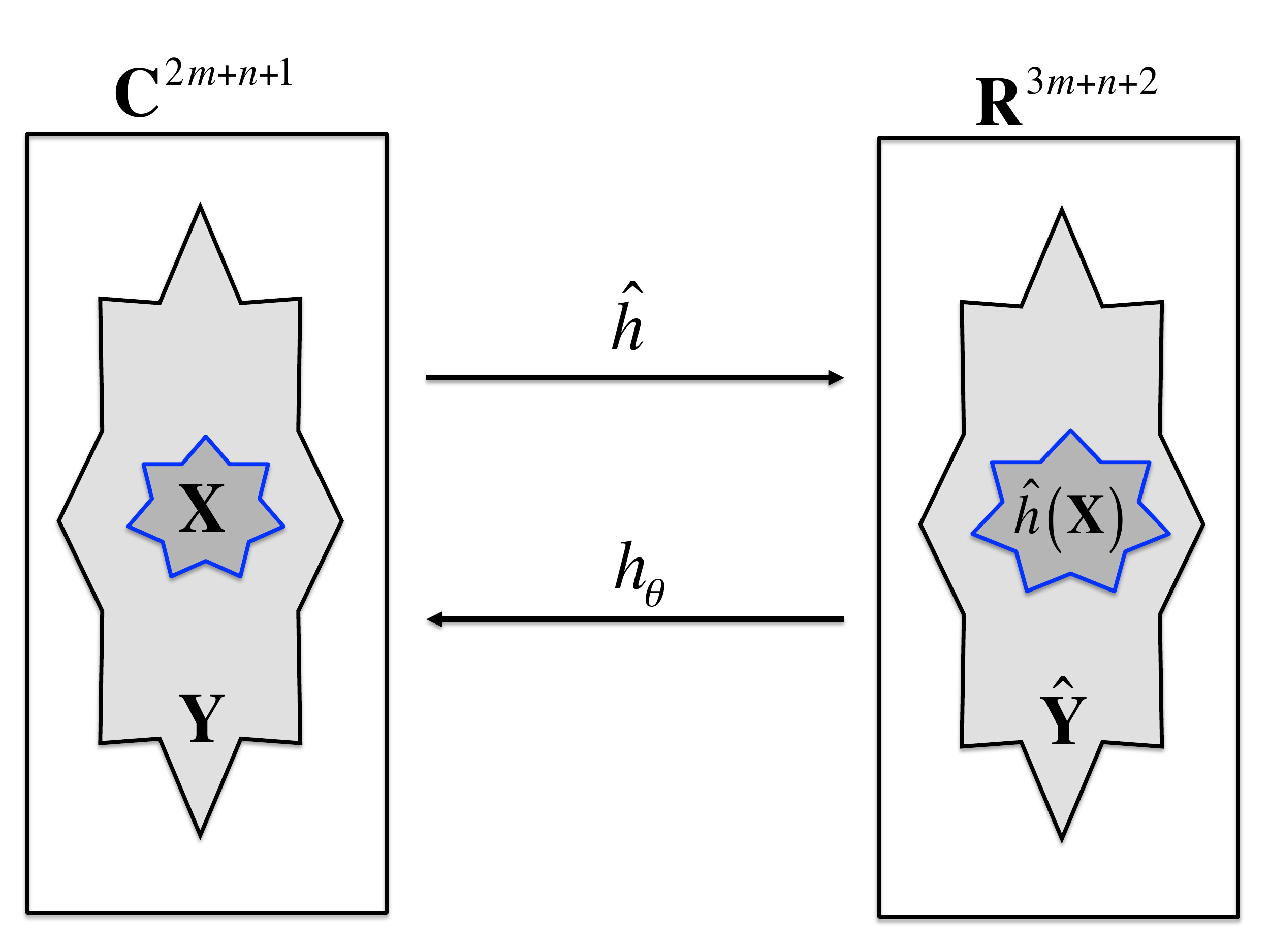}
\caption{$\mathbb X$ is the set of branch
flow solutions and $\hat{\mathbb Y} = \hat{h}(\mathbb Y)$ is the set of relaxed solutions.
The inverse projection $h_\theta$ is defined in Section V.}
\label{fig:projection}
\end{figure}
%%%%%%%%%%%%%%%%
%
Let $\mathbb{Y}\subseteq \mathbb{C}^{2m+n+1}$ denote the set of all $y := (S, I, V, s_0)$ whose
projections are the relaxed solutions:\footnote{As mentioned earlier, the set defined in
\eqref{eq:defY.1} is strictly speaking $\mathbb Y(s)$ with respect to a fixed $s$.
To simplify exposition, we abuse notation and use $\mathbb Y$ to denote both 
$\mathbb Y(s)$ and $\bigcup_{s\in \mathbb C^n} \mathbb Y(s)$, depending on the
context.   The same 
applies to $\hat{\mathbb Y}$ and $\overline{\mathbb Y}$ etc.}
\bq
\!\!\!\!\!\!\!
\mathbb{Y} & \!\! \!:= &\!\!\! \left\{ y:= (S, I, V, s_0) | \hat{h}(y) \text{ solves 
		(\ref{eq:Kirchhoff.2a})--(\ref{eq:Kirchhoff.2d})} \right\}
\label{eq:defY.1}
\eq
Define the projection
$\hat{\mathbb Y} := \hat{h}(\mathbb Y)$ of $\mathbb Y$ onto the space $\mathbb{R}^{2m+n+1}$ as
\bqn
\hat{\mathbb Y} & := & \left\{\ \hat{y}:=(S, \ell, v, s_0) \, |\,
	\hat{y} \text{ solves (\ref{eq:Kirchhoff.2a})--(\ref{eq:Kirchhoff.2d})} \ \right\}
\eqn
Clearly
\bqn
\mathbb{X}   \subseteq  \mathbb{Y} & \text{and} &
\hat{h}(\mathbb{X}) \subseteq \hat{h}(\mathbb Y) =: \hat{\mathbb{Y}}
\eqn
Their relationship is illustrated in Figure \ref{fig:projection}.

%%%%%%%%%%%%%%%%%%%%%%%%%%%%%%%%%%%%%%%%%%%%%%%%%%%
\subsection{Two relaxations}

Consider the OPF with angles relaxed:
%   \\
% \noindent
% \textbf{OPF-ar}:
\bqn
\min_{x, s} & & f \left( \hat{h}(x), s \right)
\label{eq:opf2.a'}
\\
\text{subject to} & &  x \in \mathbb Y, \quad (S, v, s_0, s) \in \mathbb S
\label{eq:opf2.c'}
\eqn
Clearly, this problem provides a lower bound to the original OPF  problem
since $\mathbb Y \supseteq \mathbb X$.   Since neither $\hat{h}(x)$ nor the constraints in $\mathbb Y$
involves angles $\angle V_i, \angle I_{ij}$, this problem is equivalent to the following
\\
 \noindent
 \textbf{OPF-ar}:
\bq
 \min_{\hat{y}, s} & & f \left( \hat{y}, s \right)
\label{eq:opf2.a}
\\
\text{subject to} & &  \hat{y} \in \hat{\mathbb Y}, \quad (S, v, s_0, s) \in \mathbb S
\label{eq:opf2.c}
\eq

The feasible set of OPF-ar is still nonconvex due to the quadratic equalities
in (\ref{eq:Kirchhoff.2d}).   Relax them to inequalities:
\bq
\ell_{ij} & \geq  & \frac{P_{ij}^2+Q_{ij}^2}{v_i}, \quad\quad (i, j) \in E
\label{eq:opf3.d}
\eq
Define the convex second-order cone (see Theorem \ref{thm:ecr} below)
 $\overline{\mathbb Y} \subseteq \mathbb{R}^{2m+n+1}$ that contains
 $\hat{\mathbb Y}$ as
\bqn
\overline{\mathbb Y} & \!\!\! := \!\!\! & \left\{ \hat{y} := (S, \ell, v, s_0) \, |\,
	\hat{y} \text{ solves (\ref{eq:Kirchhoff.2a})--(\ref{eq:Kirchhoff.2c}) and \eqref{eq:opf3.d}} 
	\right\}
\eqn
Consider the following conic relaxation of OPF-ar:
\\
\noindent
\textbf{OPF-cr}:
 \bq
\min_{\hat{y}, s} & & f \left( \hat{y}, s \right)
\label{eq:opf3.a}
\\
\text{subject to} & &  \hat{y} \in \overline{\mathbb Y}, \quad (S, v, s_0, s) \in \mathbb S
\label{eq:opf3.c}
\eq
Clearly OPF-cr provides a lower bound to OPF-ar since $\overline{\mathbb Y} \supseteq \hat{\mathbb Y}$.

%%%%%%%%%%%%%%%%%%%%%%%%%%%%%%%%%%%%%%%%%%%%%%%%%%
\subsection{Solution strategy}

In the rest of this paper, we will prove the following:
\bee
\item OFP-cr is convex.  Moreover, if there are no upper bounds on loads, then
	the conic relaxation is exact so that {\em any} optimal solution $(\hat{y}_{cr}, s_{cr})$ of OPF-cr is 
	also optimal for OPF-ar for mesh as well as radial networks (Section \ref{sec:cr}, Theorem \ref{thm:ecr}).   
	OPF-cr is a SOCP when the objective function is linear.
\item Given a solution $(\hat{y}_{ar}, s_{ar})$ of OPF-ar, if the network is radial, then we can always recover
	the phase angles $\angle V_i, \angle I_{ij}$ uniquely to obtain an optimal solution
	$(x_*, s_*)$ of the original OPF % (\ref{eq:opf1.a})--(\ref{eq:opf1.c}) 
	through an inverse projection
	(Section \ref{sec:ar}, Theorems \ref{thm.2} and \ref{thm:tree}).
\item For a mesh network, an inverse projection may not exist to map the given
	 $(\hat{y}_{ar}, s_{ar})$ to a feasible solution of OPF.   Our characterization can be used to 
	 determined if $(\hat{y}_{ar}, s_{ar})$ is globally optimal.
\eee
These results motivate the algorithm  in Figure \ref{fig:strategy}.

In Part II of this paper, we show that a mesh network can be convexified so that $(\hat{y}_{ar}, s_{ar})$ 
can always be mapped to an optimal solution of OPF for the convexified network.   
Moreover, convexification requires phase shifters only on lines outside an arbitrary spanning tree of 
the network graph. 
%	 (Section \ref{sec:mesh}, Theorem \ref{thm.3} and Corollary \ref{cor.1}).

\section{Exact conic relaxation}
\label{sec:cr}

Our first key result says that OPF-cr is exact and a SOCP
when the objective function is linear.
\begin{theorem}
\label{thm:ecr}
Suppose $\overline{p}_i^c = \overline{q}_i^c = \infty$, $i\in N$.  Then
OPF-cr  is convex.  Moreover, it is exact, i.e., {\em any} optimal solution of OPF-cr 
is also optimal for OPF-ar.
\end{theorem}
\vspace{2mm}
\begin{proof}
The feasible set is convex since the nonlinear inequalities in $\overline{\mathbb Y}$ can be written 
as the following second order cone constraint:
\begin{equation}
\begin{Vmatrix}2P_{ij}\\2Q_{ij}\\\ell_{ij}-v_i\end{Vmatrix}_2 \leq \ell_{ij} + v_i  \notag
\end{equation}
Since the objective function is convex, OPF-cr is a conic optimization.\footnote{The case of linear 
objective without line limits is proved in \cite{Farivar2011-VAR-SGC} for radial networks. 
This result is extended here to mesh networks with line limits and convex objective functions.} 
To prove that the relaxation is exact, it suffices to show that any optimal
solution of OPF-cr attains equality in (\ref{eq:opf3.d}).

Assume for the sake of contradiction that 
$(\hat{y}_*, s_*) := (S_*,  \ell_*, v_*, s_{*0}^g, s_{*0}^c, s^{g}_*, s^{c}_*)$
is optimal for OPF-cr, but  a link $(i,j)\in E$ has strict inequality, i.e.,
$[v_*]_i [\ell_*]_{ij} >  {[P_*]_{ij}}^2+{[Q_*]_{ij}}^2$.
For some $\varepsilon>0$ to be determined below, consider another point
$(\tilde{y}, \tilde{s}) =(\tilde{S}, \tilde{\ell}, \tilde{v}, \tilde{s}_0^g, \tilde{s}_0^c, \tilde{s}^g, \tilde{s}^c)$ defined by:
\[
\begin{array}{lclclcl}
\tilde{v} & = & v_*,   & &   \tilde{s}^g & = &  s_*^g   
\\
\tilde{\ell}_{ij} & = & {[\ell_*]_{ij}} - \varepsilon,   & &   \tilde{\ell}_{-ij} & = & {[\ell_*]_{-ij}}
\\
\tilde{S}_{ij} & = & {[S_*]_{ij}} - z_{ij}\varepsilon/2,  & &    \tilde{S}_{-ij} & = & [S_*]_{-ij}
\\
\tilde{s}_i^c & = & [s_*^c]_i + z_{ij} \varepsilon/2,  & &   \tilde{s}_{j}^c & = & [s_*^c]_{j} + z_{ij} \varepsilon/2
\\
\tilde{s}_{-i}^c & = & [s_*^c]_{-i},   & &   \tilde{s}_{-j}^c & = & [s_*^c]_{-j}
\end{array}
\]
where a negative index means excluding the indexed element from a vector.
Since $\tilde{\ell}_{ij}={[\ell_*]_{ij}} - \varepsilon$,
$(\tilde{y}, \tilde{s})$ has a strictly smaller objective value than $(\hat{y}_*, s_*)$
because of assumption A3.
If $(\tilde{y}, \tilde{s})$ is a feasible point, then it contradicts the optimality of $(\hat{y}_*, s_*)$.

It suffices then to check that there exists an $\varepsilon>0$ such that
$(\tilde{y}, \tilde{s})$ satisfies (\ref{GenLimits})--(\ref{eq:ll}), 
(\ref{eq:Kirchhoff.2a})--(\ref{eq:Kirchhoff.2c}) and \eqref{eq:opf3.d},
 and hence is indeed a feasible point.
Since $(\hat{y}_*, s_*)$ is feasible, (\ref{GenLimits})--\eqref{eq:ll} hold for $(\tilde{y}, \tilde{s})$ too.
Similarly, $(\tilde{y}, \tilde{s})$ satisfies (\ref{eq:Kirchhoff.2a})--(\ref{eq:Kirchhoff.2b}) at all nodes $k \neq i, j$
and (\ref{eq:Kirchhoff.2c}), \eqref{eq:opf3.d} over all links $(k, l) \neq (i, j)$.
We now show that $(\tilde{y}, \tilde{s})$ satisfies (\ref{eq:Kirchhoff.2a})--(\ref{eq:Kirchhoff.2b}) also at nodes $i, j$,
and (\ref{eq:Kirchhoff.2c}), \eqref{eq:opf3.d}  over  $(i, j)$.

Proving  (\ref{eq:Kirchhoff.2a})--(\ref{eq:Kirchhoff.2b}) is equivalent to proving (\ref{eq:Kirchhoff.1a}).  
At node $i$, we have
\bqn
\tilde{s}_{i}  & = & \tilde{s}_i^g - \tilde{s}_i^c  \ \ = \ \  {[s_*^g]_i}  - {[s_*^c]_i} - z_{ij} \varepsilon/2
 \\
 & = & \sum_{i\rightarrow j'} [S_*]_{ij'}  - \sum_{k\rightarrow i} \left( [S_*]_{ki} - z_{ki} [\ell_*]_{ki} \right)
\\
&& \ \    + y_i^* v_i -  z_{ij} \varepsilon /2
\\
& = &   \sum_{i\rightarrow j', j'\neq j} \tilde{S}_{ij'} + \left( \tilde{S}_{ij} + z_{ij}\varepsilon/2 \right)
\\
& & 
- \sum_{k\rightarrow i} \left( \tilde{S}_{ki} - z_{ki} \tilde{\ell}_{ki} \right) 
  + y_i^* \tilde{v}_i -  z_{ij} \varepsilon /2
\\  
& = & \sum_{i\rightarrow j'} \tilde{S}_{ij'} 
	- \sum_{k\rightarrow i} \left( \tilde{S}_{ki} - z_{ki} \tilde{\ell}_{ki} \right) 
	+  y_i^* \tilde{v}_i
\eqn
At node $j$, we have
\bqn
\tilde{s}_{j}  & = & \tilde{s}_j^g - \tilde{s}_j^c  
\ \ = \ \  [s_*^g]_j  - [s_*^c]_j -  z_{ij} \varepsilon/2
\\
& = &  \ \sum_{j\rightarrow k} [S_*]_{jk} - \sum_{i'\rightarrow j} \left( [S_*]_{i'j} - z_{i'j} [\ell_*]_{i'j} \right)  
\\  
&& \ + y_j^* v_j - z_{ij} \varepsilon /2 
\\ 
& = &  \sum_{j\rightarrow k} \tilde{S}_{jk} - 
\sum_{i'\rightarrow j, i'\neq i} \left( \tilde{S}_{i'j} - z_{i'j} \tilde{\ell}_{i'j} \right) + y_j^* \tilde{v}_j 
\\
& & 
- \left( (\tilde{S}_{ij} + z_{ij}\varepsilon/2) -z_{ij} (\tilde{\ell}_{ij}+\varepsilon) \right)
  -  z_{ij} \varepsilon /2
\\ 
& = & \sum_{j\rightarrow k} \tilde{S}_{jk}
	- \sum_{i'\rightarrow j} \left( \tilde{S}_{i'j} - z_{i'j} \tilde{\ell}_{i'j} \right)
		+ y_j^* \tilde{v}_j
\eqn
Hence (\ref{eq:Kirchhoff.2a})--(\ref{eq:Kirchhoff.2b}) hold at nodes $i, j$.

 For (\ref{eq:Kirchhoff.2c}) across link $(i, j)$:
\bqn
 \tilde{v}_j &=&  [v_*]_i-2(r_{ij} [P_*]_{ij}+x_{ij} [Q_*]_{ij}) 
\\
& & \quad\quad\quad\  \,  + (r_{ij}^2+x_{ij}^2) [\ell_*]_{ij}
\\
%
% &=&\tilde{v}_i-2(r_{ij} (\tilde{P}_{ij}+r_{ij}\varepsilon/2)+x_{ij}(\tilde{Q}_{ij}+x_{ij}\varepsilon/2)) 
 % \\
% && \ \ \ \ \ \ \ \ \ \ + (r_{ij}^2+x_{ij}^2) (\tilde{\ell}_{ij}+\varepsilon)\nonumber
% \\
%
&=& \tilde{v}_i-2(r_{ij}\tilde{P}_{ij}+x_{ij}\tilde{Q}_{ij}) + (r_{ij}^2+x_{ij}^2) \tilde{\ell}_{ij}
\eqn

For \eqref{eq:opf3.d}  across link $(i, j)$, we have
  \bqn
 & & \tilde{v}_i \tilde{\ell}_{ij} - \tilde{P}_{ij}^2 - \tilde{Q}_{ij}^2
 \\
  & = & {[v_*]_i}  \left( [\ell_*]_{ij} - \varepsilon \right)  - \left([P_*]_{ij}-r_{ij}\varepsilon/2\right)^2
  \\
  & & \quad\quad\quad\quad\quad\quad\ \,  -  \left([Q_*]_{ij}-x_{ij}\varepsilon/2 \right)^2
  \\
  & = &  \left(  {[v_*]_i} [\ell_*]_{ij} -  [P_*]_{ij}^2  -  [Q_*]_{ij}^2  \right)
  \\
  & & \ \ \ \
  - \varepsilon \left(  {[v_*]_i}  - r_{ij} [P_*]_{ij} -  x_{ij}[Q_*]_{ij}  \right.
  \\
  & & 
  \quad\quad\quad    \left.   + \ \varepsilon(r_{ij}^2 + x_{ij}^2)/4   \right)
\eqn
  Since ${[v_*]_i} [\ell_*]_{ij} -  [P_*]_{ij}^2  -  [Q_*]_{ij}^2 > 0$,
  we can choose an $\varepsilon>0$ sufficiently small such that
  $\tilde{\ell}_{ij}\geq  (\tilde{P}_{ij}^2+\tilde{Q}_{ij}^2)/\tilde{v}_i$. 

This completes the proof.
\end{proof}
\begin{remark}
Assumption A3 is used in the proof here to contradict the optimality of $(\hat{y}_*, s_*)$.
Instead of A3, if $f(\hat{y}, s)$ is nondecreasing in $\ell$,
the same argument shows that, given an optimal $(\hat{y}_*, s_*)$ with
a strict inequality $[v_*]_i [\ell_*]_{ij} >  {[P_*]_{ij}}^2+{[Q_*]_{ij}}^2$, one can choose
$\varepsilon>0$ to obtain another optimal point $(\tilde{y}, \tilde{s})$ that attains
equality and has a cost $f(\tilde{y}, \tilde{s}) \leq f(\hat{y}_*, s_*)$.  
Without A3, 
there is always an optimal solution of OPF-cr that is also optimal for OPF-ar, even
though it is possible that the convex relaxation OPF-cr may also have other optimal 
points with strict inequality that are infeasible for OPF-ar.
\end{remark}

\begin{remark}
\label{remark:over}
The  condition in Theorem \ref{thm:ecr} is
equivalent to the ``over-satisfaction of load'' condition in \cite{Lavaei2012, Bose2011}.
It is needed because we have increased the loads $s_*^c$ on buses $i$ and $j$ to
obtain the alternative feasible solution $(\tilde{y}, \tilde{s})$.
As we show in the simulations in \cite{Farivar-2013-BFM-TPS2}, 
it is sufficient but not necessary.
See also \cite{Gan-2012-BFMt, Li-2012-BFMt} for exact conic relaxation of OPF-cr for radial networks 
where this condition is replaced by other assumptions.
% an alternative set of assumptions.
\end{remark}

\section{Angle relaxation}
\label{sec:ar}

Theorem \ref{thm:ecr} justifies solving the convex problem OPF-cr for an
optimal solution of OPF-ar.   Given a solution $(\hat{y}, s)$ of OPF-ar,
when and how can we recover a solution $(x, s)$ of the original OPF  
(\ref{eq:opf1.a})--(\ref{eq:opf1.c})?
It depends on whether we can recover a solution $x$
to the branch flow equations \eqref{eq:Kirchhoff.1b}--\eqref{eq:Kirchhoff.1a}
from $\hat{y}$, given any $s$.    

Hence,  for the rest of Section \ref{sec:ar}, we fix an $s$.
We abuse notation in this section and write $x, \hat{y}, \theta, \mathbb X, \mathbb Y,
\hat{\mathbb Y}$ instead of $x(s), \hat{y}(s), \theta(s), \mathbb X(s), \mathbb Y(s), 
\hat{\mathbb Y}(s)$ respectively.

%%%%%%%%%%%%%%%%%%%%%%%%%%%%%%%%%%%%%%%%%%%%%%%%%%%%
\subsection{Angle recovery condition}
\label{subsec:arc}

Fix a relaxed solution $\hat{y} := (S, \ell, v, s_0) \in \hat{\mathbb Y}$.
Define the $(n+1) \times m$ incidence matrix $C$ of $G$ by
\bq
C_{ie} & = & \begin{cases}
		1 & \text{ if link $e$ leaves node $i$} \\
		-1 & \text{ if link $e$ enters node $i$} \\
		0 & \text{ otherwise}
		\end{cases}
		\quad\quad  
\label{eq:defC}
\eq
The first row of $C$ corresponds to node $0$ where $V_0 = |V_0| e^{\ii \theta_0}$ is given. 
In this paper we will only work with the $m \times n$ {\em reduced} incidence matrix $B$
obtained from $C$ by removing the first row (corresponding to $V_0$) and
taking the transpose, i.e., for $e \in E, i = 1, \dots, n$, 
\bqn
B_{ei} & = & \begin{cases}
		1 & \text{ if link $e$ leaves node $i$} \\
		-1 & \text{ if link $e$ enters node $i$} \\
		0 & \text{ otherwise}
		\end{cases},
		\quad\quad  
\eqn
Since $G$ is connected,  $m \geq n$ and rank$(B)=n$  \cite{Foulds1992}.
Fix any spanning tree $T = (N, E_T)$ of $G$.   We can assume without loss of generality 
(possibly after re-labeling some of the links) that $E_T$ consists of links $e = 1, \dots, n$.  
Then $B$ can be partitioned into
\bq
B & = & \begin{bmatrix}
		B_T  \\ B_{\perp}
		\end{bmatrix}
\label{eq:B}
\eq
where the $n\times n$ submatrix $B_T$ corresponds to links in $T$ and the 
$(m-n) \times n$ submatrix $B_{\perp}$ corresponds to links in $T^\perp := G\setminus T$.

Let $\beta := \beta(\hat{y}) \in (-\pi, \pi]^m$ be defined by:
\bq
\beta_{ij} & := & \angle \left( v_i - z_{ij}^* S_{ij} \right), \quad\quad (i, j)\in E
\label{eq:defb.1}
\eq
Informally, $\beta_{ij}$ is the phase angle difference across link $(i,j)$ that is implied by
the relaxed solution $\hat{y}$.
Write $\beta$ as
\bq
\beta & = & \begin{bmatrix}
		\beta_T  \\  \beta_{\perp}
		\end{bmatrix}
\label{eq:defb.2}
\eq 
where $\beta_T$ is $n\times 1$ and $\beta_{\perp}$ is $(m-n) \times 1$.

Recall the projection mapping 
 $\hat{h}: \mathbb{C}^{2m+n+1} \rightarrow \mathbb{R}^{3m+n+2}$ 
defined in (\ref{eq:defhhat.1})--\eqref{eq:defhhat.3}.  
For each $\theta := (\theta_i, i = 1, \dots, n) \in (-\pi, \pi]^n$,  define the inverse projection
 $h_\theta: \mathbb{R}^{3m+n+2} \rightarrow \mathbb{C}^{2m+n+1}$ by
$h_\theta(P, Q, \ell, v, p_0, q_0) = (S, I, V, s_0)$ where
\bq
S_{ij} & := & P_{ij} + \ii Q_{ij}
\label{eq:bfs.1}
\\
I_{ij} & := & \sqrt{\ell_{ij}}\ e^{\ii (\theta_i - \angle S_{ij})}
\label{eq:bfs.3}
\\
V_i & := & \sqrt{v_i} \ e^{\ii \theta_i}
\label{eq:bfs.2}
\\
s_0 & := & p_0 + \ii q_0
\label{eq:bfs.4}
\eq
These mappings are illustrated in Figure \ref{fig:projection}.

By definition of $\hat{h}(\mathbb X)$ and $\hat{\mathbb Y}$, a branch flow solution 
in $\mathbb X$  can be recovered from a given relaxed solution $\hat{y}$ 
if $\hat{y}$ is in $\hat{h}(\mathbb X)$ and cannot if  
$\hat{y}$ is in $\hat{\mathbb Y} \setminus \hat{h}(\mathbb X)$.  
In other words, $\hat{h}(\mathbb{X})$ consists of exactly  those points $\hat{y} \in \hat{\mathbb{Y}}$ for 
which there exist $\theta$ such that their inverse projections $h_\theta(\hat{y})$ are in $\mathbb X$. 
Our next key result characterizes the exact condition under which such an inverse projection
exists, and provides an explicit expression for
recovering the phase angles $\angle V_i, \angle I_{ij}$ from the given $\hat{y}$.

A {\em cycle} $c$ in $G$ is an {\em ordered} list $ c = (i_1, \dots, i_k)$ of nodes in $N$ such that 
$(i_1\sim i_2), \dots, (i_k \sim i_1)$ are all links in $E$.
We will use `$(i,j)\in c$' to denote a link $i\sim j$ in the cycle $c$.
Each  link $i\sim j$ may be in the same
orientation $((i,j)\in E)$ or in the opposite orientation $((j,i)\in E)$.
Let $\tilde{\beta}$ be the extension of $\beta$ from
directed links to undirected links: if $(i,j)\in E$ then $\tilde{\beta}_{ij} := \beta_{ij}$ and 
$\tilde{\beta}_{ji} := - \beta_{ij}$.
% Suppose each cycle in $G$ is endowed with an arbitrary orientation (clockwise or counterclockwise).
% Given a cycle $c$ and a link $e\in c$, let $\delta (e, c) = 1$ if $e$ has the same orientation as $c$
% and $-1$ otherwise. 
For any $d$-dimensional vector $\alpha$, let $\mathcal{P}(\alpha)$ denote its projection
onto $(-\pi, \pi]^d$ by taking modulo $2\pi$ componentwise.
\begin{theorem}
\label{thm.2}
Let $T$ be {\em any} spanning tree of $G$.
Consider a relaxed solution $\hat{y} \in \hat{\mathbb Y}$
and the corresponding $\beta = \beta(\hat{y})$ defined in
(\ref{eq:defb.1})--\eqref{eq:defb.2}.
\bee
\item There exists a unique $\theta_* \in (-\pi, \pi]^n$ such that $h_{\theta_*}(\hat{y})$ is a branch 
	flow solution in $\mathbb X$  if and only if 
	\bq
	B_{\perp} B_T^{-1} \beta_T & = & \beta_{\perp}  \ \ \ ( \text{mod } 2\pi )
	\label{eq:recoverycond}
	\eq
	
\item The angle recovery condition \eqref{eq:recoverycond} holds if and only if for every cycle $c$
	in $G$
	\bq
	\sum_{(i,j) \in c} \tilde{\beta}_{ij} = 0 \ \ \  ( \text{mod } 2\pi )
	\label{eq:cyclecond}
	\eq
	
\item If \eqref{eq:recoverycond} holds then $\theta_* = \mathcal{P}\left(B_T^{-1}\beta_T\right)$.
\eee
\end{theorem}

\begin{remark}
Given a relaxed solution $\hat{y}$, Theorem \ref{thm.2} prescribes a 
way to check if a branch
flow solution can be recovered from it, and if so, the required computation.
The angle recovery condition \eqref{eq:recoverycond}  % is a condition on $\hat{y}$ and
depends only on the network topology through the reduced incidence matrix $B$.
The choice of spanning tree $T$ corresponds to choosing $n$ linearly independent 
rows of $B$ to form $B_T$ and does not affect the conclusion of the theorem.
\end{remark}

\begin{remark}
\label{remark:cycle}
When it holds, the angle recovery condition \eqref{eq:cyclecond} has a familiar interpretation
(due to Lemma \ref{thm.1} below):  the voltage angle differences (implied by $\hat{y}$) 
sum to zero (mod $2\pi$) around any cycle.
\end{remark}

\begin{remark}
\label{remark:eq}
A direct consequence of Theorem \ref{thm.2} is that the relaxed branch flow model 
\eqref{eq:Kirchhoff.2a}--\eqref{eq:Kirchhoff.2d} together with the angle recovery condition 
\eqref{eq:recoverycond} is equivalent to the original branch flow model 
\eqref{eq:Kirchhoff.1b}--\eqref{eq:Kirchhoff.1a}.
That is, {\em $x$ satisfies \eqref{eq:Kirchhoff.1b}--\eqref{eq:Kirchhoff.1a} if and only if
$\hat{y} = \hat{h}(x)$ satisfies \eqref{eq:Kirchhoff.2a}--\eqref{eq:Kirchhoff.2d} and
\eqref{eq:recoverycond}.}
The challenge in computing a branch flow solution $x$ is that \eqref{eq:recoverycond}
is nonconvex.
\end{remark}
\vspace{0.1in}

The proof of Theorem \ref{thm.2} relies on the following important lemma that gives a
necessary and sufficient condition for 
an inverse projection $h_\theta(\hat{y})$ defined by \eqref{eq:bfs.1}--\eqref{eq:bfs.4}
to be a branch flow solution in $\mathbb X$.
Fix  any $\hat{y} := (S, \ell, v, s_0)$ in $\hat{\mathbb Y}$ and the corresponding 
$\beta := \beta(\hat{y})$ defined in (\ref{eq:defb.1}).  Consider the equation
	\bq
	\label{eq:theta}
	B\theta & = & \beta + 2\pi k
	\eq
where $k \in \mathbb N^m$ is an integer vector.  
Since $G$ is connected,  $m \geq n$ and rank$(B)=n$.  Hence, given any $k$,
there is at most one $\theta$ that solves \eqref{eq:theta}.  Obviously, given any
$\theta$, there is exactly one $k$ that solves \eqref{eq:theta}; we denote it 
by $k(\theta)$ when we want to emphasize the dependence on $\theta$.
Given any solution $(\theta, k)$ with $\theta \in (-\pi, \pi]^n$, define its 
{\em equivalence class} by \footnote{Using the connectedness of $G$ and
the definition of $B$, one can argue that $\alpha$ must be an
integer vector for $k+B\alpha$ to be integral.}
\bqn
\sigma(\theta, k) & := & \{ (\theta + 2\pi \alpha, k+ B\alpha) \ |\ \alpha \in \mathbb{N}^n \}
\eqn
We say {\em $\sigma(\theta, k)$ is a solution of \eqref{eq:theta}} if every vector in
$\sigma(\theta, k)$ is a solution of \eqref{eq:theta}, and  {\em $\sigma(\theta, k)$ is 
the unique solution of \eqref{eq:theta}} if it is the only equivalence class of solutions.
\begin{lemma}
\label{thm.1}
Given any $\hat{y} := (S, \ell, v, s_0)$ in $\hat{\mathbb Y}$ and the corresponding 
$\beta := \beta(\hat{y})$ defined in (\ref{eq:defb.1}):
\bee
\item $h_{\theta}(\hat{y})$ is a branch flow solution in $\mathbb X$ if and only if
	$(\theta, k(\theta))$ solves \eqref{eq:theta}.

\item there is at most one $\sigma(\theta, k)$, $\theta\in (-\pi, \pi]^n$, that is the
	unique solution of \eqref{eq:theta}, when it exists.
\eee
\end{lemma}
\begin{proof}
Suppose $(\theta, k)$ is a solution of \eqref{eq:theta} for some $k = k(\theta)$.
We need to show that  (\ref{eq:Kirchhoff.2a})--(\ref{eq:Kirchhoff.2d}) together with
\eqref{eq:bfs.1}--\eqref{eq:bfs.4} and \eqref{eq:theta} imply (\ref{eq:Kirchhoff.1b})--(\ref{eq:Kirchhoff.1a}).
Now (\ref{eq:Kirchhoff.2a}) and (\ref{eq:Kirchhoff.2b}) are equivalent to (\ref{eq:Kirchhoff.1a}).  Moreover
(\ref{eq:Kirchhoff.2d}) and \eqref{eq:bfs.1}--\eqref{eq:bfs.2} imply \eqref{eq:Kirchhoff.1c}.
To prove \eqref{eq:Kirchhoff.1b}, substitute \eqref{eq:Kirchhoff.1c} into \eqref{eq:theta} to get
\bqn
\theta_i - \theta_j & = & \angle \left( v_i - z_{ij}^* V_i I_{ij}^* \right) + 2\pi k_{ij}
\\
& = &  \angle \ V_i \left(V_i - z_{ij} I_{ij} \right)^* + 2\pi k_{ij}
\eqn
Hence
\bq
\angle V_j  & = & \theta_j \ = \ \angle \left(V_i - z_{ij} I_{ij}\right) - 2\pi k_{ij}
\label{eq:aVj}
\eq
From \eqref{eq:Kirchhoff.2c} and (\ref{eq:Kirchhoff.1c}), we have
\bqn
|V_j|^2 & = & |V_i|^2 + |z_{ij}|^2 |I_{ij}|^2 - (z_{ij} S_{ij}^* + z_{ij}^* S_{ij})
\\
	& = & |V_i|^2 + |z_{ij}|^2 |I_{ij}|^2 - (z_{ij} V_i^* I_{ij} + z_{ij}^* V_i I_{ij}^*)
\\
	& = & |V_i - z_{ij} I_{ij}|^2
\eqn
This and (\ref{eq:aVj}) imply $V_j = V_i - z_{ij} I_{ij}$ which is (\ref{eq:Kirchhoff.1b}).

Conversely, suppose $h_{\theta}(\hat{y}) \in \mathbb X$.  
From \eqref{eq:Kirchhoff.1b} and \eqref{eq:Kirchhoff.1c}, we have
$V_i V_j^* = |V_i|^2 - z_{ij}^* S_{ij}$.   Then 
${\theta}_i - {\theta}_j = \beta_{ij} + 2\pi k_{ij}$
for some integer $k_{ij} = k_{ij}(\theta)$.   Hence $(\theta, k)$ solves \eqref{eq:theta}.

The discussion preceding the lemma shows that, given any $k\in \mathbb N^m$,
there is at most one $\theta$ that satisfies \eqref{eq:theta}.   If no such $\theta$ exists for any
$k\in\mathbb N^m$, then \eqref{eq:theta} has no solution $(\theta, k)$.  If \eqref{eq:theta} has
a solution $(\theta, k)$, then clearly $(\theta + 2\pi \alpha, k+ B\alpha)$ are also solutions
for all $\alpha\in \mathbb{N}^n$.
% (i.e., adding $2\pi$ to $\theta_i$ and an integer vector in the 
% range space of $B$ to $k$ gives another solution). 
Hence we can assume without loss of generality that $\theta\in (-\pi, \pi]^n$.  
We claim that $\sigma(\theta, k)$ is the unique solution of \eqref{eq:theta}.
Otherwise, there is an $(\tilde{\theta}, \tilde{k})\not\in \sigma(\theta, k)$ with
$B\tilde{\theta} = \beta + 2\pi \tilde{k}$.  
Then $B(\tilde{\theta} - \theta) = 2\pi (\tilde{k} - k)$,
or $\tilde{k} = k + B\alpha$ for some $\alpha$.  Since $\tilde{k}\in \mathbb N^m$, $\alpha$ is
an integer vector; moreover $\tilde{\theta}$ is unique given $\tilde{k}$.  This means 
$(\tilde{\theta}, \tilde{k}) \in \sigma(\theta, k)$, a contradiction. 
\end{proof}

%%%%%%%%%%%%%%%%%%%%%%%%%%%%%%%%%%%%%%%%%%%%%%%%%%%%
\vspace{0.1in}
\begin{proof}[Proof of Theorem \ref{thm.2}]
Since $m \geq n$ and rank$(B)=n$, 
we can always find $n$ linearly independent rows of $B$ to form
a basis.    The choice of this basis corresponds to choosing a { spanning} tree of
$G$, which always exists since $G$ is connected \cite[Chapter 5]{Biggs-1993-agt}.
Assume without loss of generality that the first $n$ rows is such a basis so that $B$
and $\beta$ are partitioned as in \eqref{eq:B} and \eqref{eq:defb.2} respectively.
Then Lemma \ref{thm.1} implies that $h_{\theta_*}(\hat{y}) \in \mathbb X$ with
$\theta_*\in (-\pi, \pi]^n$ if and only if $(\theta_*, k_*(\theta_*))$ is the unique solution
of
\bq
\begin{bmatrix}
B_T \\ B_{\perp}
\end{bmatrix}
\theta & = &
\begin{bmatrix}
\beta_T \\ \beta_{\perp}
\end{bmatrix}
+ 2\pi
\begin{bmatrix}
k_T \\ k_{\perp}
\end{bmatrix}
\label{eq:atb.2}
\eq
Since $T$ is a spanning tree, the $n\times n$ submatrix $B_T$ 
is invertible.  
Moreover \eqref{eq:atb.2} has a unique solution if and only if 
$B_{\perp} B_T^{-1} (\beta_T + 2\pi k_T) = \beta_{\perp} + 2\pi k_\perp$, 
i.e., $B_\perp B_T^{-1} \beta_T = \beta_\perp + 2\pi \hat{k}_\perp$
where $\hat{k}_\perp := k_\perp - B_\perp B_T^{-1}k_T$.   Then
 \eqref{eq:Binv} below implies that $\hat{k}_\perp$ is an integer vector.
This proves the first assertion.

For the second assertion, 
recall that the spanning tree $T$ defines the orientation of all links in $T$ to be directed
away from the root node $0$.  
  Let $T(i\leadsto j)$ denote the unique path from node $i$ to node $j$ in $T$;
in particular, $T(0 \leadsto j)$ consists of links all with the same orientation as the path and 
$T(j \leadsto 0)$ of links all with the opposite orientation.
Then it can be verified directly that 
\bq
\left[ B_T^{-1} \right]_{ei} & \!\!\! := \!\!\! & \begin{cases}
		-1  &  \text{ if link $e$ is in $T(0 \leadsto i)$}
		\\
		0   &  \text{ otherwise}
		\end{cases}
\label{eq:Binv}
\eq 
Hence $B_T^{-1} \beta_T$ represents the (negative of the) sum of angle differences on the
path $T(0\leadsto i)$ for each node $i\in T$:
\bqn
\left[ B_T^{-1} \beta_T \right]_i & = & \sum_e \left[B_T^{-1} \right]_{ie} \left[ \beta_T \right]_e
	\ = \ - \sum_{e \in T(0\leadsto i)} \left[ \beta_T \right]_e
\eqn
Hence $B_\perp B_T^{-1} \beta_T$ is the sum of voltage angle differences
from node $i$ to  node $j$ along the unique path in $T$, for every link $(i,j) \in E\setminus E_T$
not in the tree $T$.
To see this, we have, for each link $e := (i,j) \in E\setminus E_T$, 
\bqn
\left[ B_\perp B_T^{-1} \beta_T \right]_e & = & \left[ B_T^{-1} \beta_T \right]_i - \left[ B_T^{-1} \beta_T \right]_j
\\
& = &  \sum_{e' \in T(0\leadsto j)} \left[ \beta_T \right]_{e'}  - \sum_{e' \in T(0\leadsto i)} \left[ \beta_T \right]_{e'}
\eqn
Since
\bqn
\sum_{e' \in T(0\leadsto j)} \left[ \beta_T \right]_{e'} & = &  - \sum_{e' \in T(j\leadsto 0)} \left[ \tilde{\beta}_T \right]_{e'}
\eqn
the angle recovery condition \eqref{eq:recoverycond} is equivalent to
\bqn
&   &  \sum_{e' \in T(0\leadsto i)} \left[ \beta_T \right]_{e'} + \left[ \beta_\perp \right]_{ij} + 
\sum_{e' \in T(j \leadsto 0)} \left[ \tilde{\beta}_T \right]_{e'}  
\\
& = & \sum_{e'\in c(i,j)} \tilde{\beta}_{e'} \ \ = \ \ 0 \ \ \ (\text{mod } 2\pi )
\eqn
where $c(i,j)$ denotes the unique basis cycle (with respect to $T$) associated with each link  $(i,j)$ 
not in $T$ \cite[Chapter 5]{Biggs-1993-agt}.   Hence \eqref{eq:recoverycond} is equivalent to (\ref{eq:cyclecond}) 
on all basis cycles, and therefore it is equivalent to (\ref{eq:cyclecond}) on all cycles.

Suppose \eqref{eq:recoverycond} holds and let $(\theta_*, k_*)$ be the unique solution 
of \eqref{eq:atb.2} with $\theta_* \in (-\pi, \pi]^n$.   
We are left to show that  $\theta_* = \mathcal{P}\left(B_T^{-1}\beta_T\right)$. 
By \eqref{eq:atb.2} we have $\theta_* - 2\pi B_T^{-1} [k_*]_T = \beta_T$.
Consider $\alpha := - B_T^{-1} [k_*]_T$ which is in $\mathbb N^n$ due to \eqref{eq:Binv}.
Then $(\theta_* + 2\pi \alpha, k_* + B\alpha)\in \sigma(\theta_*, k_*)$ and hence is also a solution
of \eqref{eq:atb.2} by Lemma \ref{thm.1}.  
% Moreover the proof of Lemma \ref{thm.1} shows that, given 
% $k_* + B\alpha$, $\theta_* + 2\pi \alpha$ is the only solution of \eqref{eq:atb.2}.
Moreover $\theta_* + 2\pi \alpha = B_T^{-1} \beta_T$
since $[k_*]_T + B_T\alpha = 0$.   This means that 
$\theta_*$ is given by $\mathcal{P}\left(B_T^{-1}\beta_T\right)$ since $\theta_*\in (-\pi, \pi]^n$.
\end{proof}
%%%%%%%%%%%%%%%%%%%%%%%%%%%%%%%%%%%%%%%%%%%%%%%%%%%%

%%%%%%%%%%%%%%%%%%%%%%%%%%%%%%%%%%%%%%%%%%%%%%%%%%%%
\subsection{Angle recovery algorithms}
\label{subsec:ara}

Theorem \ref{thm.2} suggests a centralized method to compute a branch flow solution from a
relaxed solution.
 \\
 \noindent
 \textbf{Algorithm 1: centralized angle recovery.}
 Given a relaxed solution $\hat{y}  \in \hat{\mathbb Y}$,
 \bee
 \item Choose {\em any} $n$ basis rows of $B$ and form $B_T$, $B_{\perp}$.
 \item Compute $\beta$ from $\hat{y}$ and check if $B_{\perp} B_T^{-1} \beta_T - \beta_{\perp} = 0$
 	(mod $2\pi$).
 \item If not, then  $\hat{y} \not\in \hat{h}(\mathbb X)$;  stop.
 \item Otherwise,  compute $\theta_* = \mathcal{P} \left( B_T^{-1} \beta_T \right)$.
\item Compute $h_{\theta_*}(\hat{y}) \in \mathbb X$ through \eqref{eq:bfs.1}--\eqref{eq:bfs.4}.
\eee
Theorem \ref{thm.2} guarantees that $h_{\theta_*}(\hat{y})$, if exists, is the unique branch flow
solution of (\ref{eq:Kirchhoff.1b})--(\ref{eq:Kirchhoff.1a})  whose projection is
$\hat{y}$.

The relations (\ref{eq:Kirchhoff.1c}) and (\ref{eq:theta}) motivate an alternative procedure to compute the
angles $\angle I_{ij}$, $\angle V_i$, and  a branch flow solution.
This procedure is more amenable to a distributed implementation.  
% Starting from the root node 0, for all children $j\in\delta(0)$, 
% use (\ref{eq:Kirchhoff.1c}) to determine 
% \bqn
% \angle I_{0j} & := & \angle V_0 - \angle S_{0j}
% \eqn
% and use (\ref{eq:theta}) to determine
% \bqn
% \angle V_j & := & \angle V_0 - \angle (v_0 - z_{0j}^*S_{0j})
% \eqn
%  Repeat this process, starting from $V_j$, $j\in \delta(0)$, to determine the
% angles $\angle V_k, \angle I_{jk}$ for all children $k\in\delta(j)$, and so on, until all
% nodes in the radial network are covered.   In summary: 
\\
\noindent
\textbf{Algorithm 2: distributed angle recovery.}
Given a relaxed solution $\hat{y}  \in \hat{\mathbb Y}$,
\bee
\item Choose {\em any} spanning tree $T$ of $G$ rooted at node 0.

\item For $j = 0, 1, \dots, n$ (i.e., as  $j$ ranges over the tree $T$, starting from the
	root and in the order of breadth-first search), for all children $k$ with $j\rightarrow k$, set
\bq
\angle I_{jk} & := & \angle V_j - \angle S_{jk}
\label{eq:phase.1}
\\
\angle V_k & := & \angle V_j - \angle (v_j - z_{jk}^*S_{jk})
\label{eq:phase.2}
\eq

\item For each link $(j, k)\in E\setminus E_T$ not in the
	spanning tree, node $j$ is an additional parent of $k$ in addition to $k$'s parent in the spanning
	tree from which $\angle V_k$ has already been computed in Step 2.   
	\bee
	\item Compute current angle $\angle I_{jk}$ using (\ref{eq:phase.1}).
	\item Compute a new voltage angle 	$\theta_k^j$ using the new parent $j$ and (\ref{eq:phase.2}).
	If $\theta_k^j \angle V_k \neq 0$ (mod $2\pi$), then angle recovery has failed; stop.
	\eee
	
\eee
If the angle recovery procedure succeeds in Step 3,
then $\hat{y}$ together with these angles $\angle V_k, \angle I_{jk}$
are indeed a branch flow solution.
Otherwise, a link $(j,k)$ not in the tree $T$ has been identified where condition \eqref{eq:cyclecond}
is violated over the unique basis cycle (with respect to $T$) associated with link $(j,k)$.

%%%%%%%%%%%%%%%%%%%%%%%%%%%%%%%%%%%%%%%%%%%%%%%%%%%
\subsection{Radial networks}

Recall that all relaxed solutions in
$\hat{\mathbb Y} \setminus \hat{h}(\mathbb X)$ are spurious.   Our next key result shows that,
for radial network, $\hat{h}(\mathbb X) = \hat{\mathbb Y}$ and hence angle relaxation is 
always exact in the sense that  there is always a unique inverse projection that maps any relaxed 
solution $\hat{y}$ to a branch flow solution in $\mathbb X$ (even though $\mathbb X \neq \mathbb Y$).
\begin{theorem}
\label{thm:tree}
Suppose $G=T$ is a tree.   Then  
\bee
\item $\hat{h}(\mathbb X) = \hat{\mathbb Y}$.

\item given any $\hat{y}$, $\theta_* := \mathcal{P} \left( B^{-1}\beta \right)$ always exists and is 
	the unique vector in $(-\pi, \pi]^n$ such that $h_{\theta_*}(\hat{y}) \in \mathbb X$.
\eee
\end{theorem}
\begin{proof}
When $G = T$ is a tree, $m=n$ and hence $B = B_T$ and $\beta = \beta_T$.
Moreover $B$ is $n \times n$ and of full rank.
Therefore $\theta_* =  \mathcal{P} \left( B^{-1}\beta \right) \in (\pi, \pi]^n$ always exists and, 
by Theorem \ref{thm.2},
$h_{\theta_*}(\hat{y})$ is the unique branch flow solution in $\mathbb X$ whose
projection is $\hat{y}$.  
Since this holds for any arbitrary $\hat{y} \in \hat{\mathbb Y}$, $\hat{\mathbb Y} = \hat{h}(\mathbb X)$.
\end{proof}

\vspace{3mm}
A direct consequence of Theorem \ref{thm:ecr} and Theorem \ref{thm:tree} is that, for a radial network,
OPF is equivalent to the convex problem OPF-cr in the sense that we can obtain an optimal solution 
of one problem from that of the other.
\begin{corollary}
\label{corollary:tree}
Suppose $G$ is a tree.
Given {\em any} optimal solution $(\hat{y}_*, s_*)$ of OPF-cr,
	there exists a unique $\theta_* \in (-\pi, \pi]^n$ such that $(h_{\theta_*}(\hat{y}_*), s_*)$ 
	is  optimal for  OPF.
\end{corollary}
% \begin{proof}
% Suppose $(\hat{y}_*, s_*)$ is optimal for OPF-cr  (\ref{eq:opf3.a})--(\ref{eq:opf3.c}).
% Theorem \ref{thm:ecr} implies that it is also optimal for OPF-ar.  In particular $\hat{y}_* \in \hat{\mathbb Y}(s_*)$.
% Since $G$ is a tree, $\hat{\mathbb Y}(s_*) = \hat{h}(\mathbb X(s_*))$ by Theorem \ref{thm:tree} and hence there is
% a unique $\theta_*$ such that $h_{\theta_*}(\hat{y}_*)$ is a branch flow solution in $\mathbb X(s_*)$.
% This means $(h_{\theta_*}(\hat{y}_*), s_*)$ is feasible for 
% OPF (\ref{eq:opf1.a})--(\ref{eq:opf1.c}).   Since OPF-ar is a relaxation of OPF, $(h_{\theta_*}(\hat{y}_*), s_*)$
% is also optimal for OPF.
% \end{proof}

\section{Conclusion}
\label{sec:c}

We have presented a branch flow model  for the 
analysis and optimization of mesh as well as radial networks.    
We have proposed a solution strategy for OPF that consists of two steps:
\bee
\item Compute a relaxed solution of OPF-ar by solving its second-order conic relaxation OPF-cr.
\item Recover from a relaxed solution an optimal solution of the original OPF using an angle 
	recovery algorithm, if possible.
\eee
We have proved that this strategy guarantees a globally optimal solution
for radial networks, provided there are
no upper bounds on loads.   For mesh networks the angle recovery condition may not hold
but can be used to check if a given relaxed solution is globally optimal.

The branch flow model is an alternative  to the bus injection model.  
% It adds to our
% understanding of power flow solutions and their relaxations, especially with the 
% clarification of the equivalence of these two models \cite{Bose-2012-BFMe-Allerton}.
It has the advantage that its variables correspond directly to physical quantities, such
as branch power and current flows, and therefore are often more intuitive than a semidefinite
matrix in the bus injection model.
For instance, Theorem \ref{thm.2} implies
that the number of power flow solutions depends only on the magnitude of 
voltages and currents, not on their phase angles.

\section*{Acknowledgment}
We are grateful to S.  Bose, K. M. Chandy and L. Gan of Caltech,
 C.  Clarke, M. Montoya, and R.  Sherick
of the Southern California Edison (SCE), and B. Lesieutre of Wisconsin
 for helpful discussions.   We acknowledge the support of
NSF through NetSE grant CNS 0911041, DoE's
ARPA-E through grant DE-AR0000226,  the National Science Council
of Taiwan (R. O. C.) through grant NSC 101-3113-P-008-001,
SCE, 
the Resnick Institute of Caltech, 
Cisco, and the Okawa Foundation.
% Jeff Gooding, Russel Neal, Percy Haralson, Michael Montoya, Bryan Pham, Robert Yinger and Marshall Parsons in the Advanced Technology Division of
%the Southern California Edison's (SCE) Transmission and Distribution Business Unit (TDBU) for multiple fruitful discussions.
%

% \bibliographystyle{unsrt}
%\bibliography{../../../PowerRef-201202.bib}
%

\vspace{0.2in}
\noindent
\textbf{See Part II of this paper \cite{Farivar-2013-BFM-TPS2} for author biographies.} 

\end{document}

\title{Branch Flow Model: Relaxations and Convexification  {\Large (Part II)}
\thanks{\textbf{To appear in {\em IEEE Trans. Power Systems}, 2013}
(submitted in May 11, 2012, accepted for publication on March 3, 2013).
A preliminary and abridged version has appeared in \cite{Farivar-2012-BFM-CDC}.}
}

\author{
Masoud Farivar \quad \quad \quad
Steven H. Low \\
Engineering and Applied Science  \\
Caltech
}
\maketitle

\begin{abstract}
We propose a branch flow model for the analysis and optimization of mesh
as well as radial networks.   The model leads to a new approach to solving optimal
power flow (OPF) that consists of two relaxation steps.   The first step
eliminates the voltage and current angles and the second step approximates the
resulting problem by a conic program that can be solved efficiently.
For radial networks, we prove that both relaxation steps are always exact, provided
there are no upper bounds on loads.   For mesh networks, the conic relaxation is
always exact but the angle relaxation may not be exact,
and we provide a simple way to determine if a relaxed solution is globally optimal.
We propose convexification of mesh networks using phase shifters so
that  OPF for the convexified network can
always be solved efficiently for an optimal solution.   We prove that convexification
requires phase shifters only outside a spanning tree of the network
and their placement depends only on network topology, not on power flows,
generation, loads, or operating constraints.
Part I  introduces our branch flow model, explains the two relaxation steps,
and proves the conditions for exact relaxation.
Part II describes convexification of mesh networks, and presents simulation
results.

\end{abstract}
%

% \newpage
% \tableofcontents
% \newpage

%%-------------------------------------------------------------------
% \input{intro-2.tex}
% \input{mesh.tex}
% \input{Simulations.tex}
% \input{conc-2.tex}
%%-------------------------------------------------------------------

\section{Introduction}

In Part I of this two-part paper \cite{Farivar-2013-BFM-TPS1},
we introduce a branch flow model that focuses on branch 
variables instead of nodal variables.   We formulate optimal power flow (OPF) within the
branch flow model and propose two relaxation steps.  The first step eliminates phase
angles of voltages and currents.
We call the resulting problem OPF-ar which is still nonconvex.
The second step relaxes the feasible set of OPF-ar to a 
second-order cone.  We call the resulting problem OPF-cr which is convex,
indeed a second-order cone
program (SOCP) when the objective function is linear.
We prove that the conic relaxation OPF-cr is always exact  {\em even} for
mesh networks, provided there are no upper bounds on real and reactive loads, i.e., 
{ any} optimal solution of OPF-cr is also optimal for OPF-ar.  
Given an optimal solution of OPF-ar, whether we can derive  an optimal solution to the original OPF
depends on whether we can recover the voltage and current angles correctly from the given OPF-ar
solution.  
We characterize the exact condition (the angle recovery condition)
under which this is possible, and present two angle recovery algorithms.
It turns out that the angle recovery condition has a simple interpretation: any  solution
of OPF-ar implies a phase angle difference across a line, and the angle recovery condition
says that the implied phase angle differences sum to zero (mod $2\pi$) around each cycle.
For a radial network, this condition holds trivially and hence 
solving the conic relaxation OPF-cr always produces an optimal solution for the original OPF.
For a mesh network, the angle recovery condition may not hold,  and our characterization 
can be used to check if a relaxed solution yields an optimal solution for OPF.

In this paper, we prove that, by placing phase shifters on some of the branches, 
{\em any} relaxed solution of OPF-ar 
can be mapped to an optimal solution of OPF for the convexified network, with an optimal cost
that is no higher than that of the original network.   Phase shifters thus convert an NP-hard
problem into a simpler problem.   Our result implies that when the angle
recovery condition holds for a relaxed branch flow solution, not only is the solution optimal for
the OPF without phase shifters, but the addition of phase shifters cannot further
reduce the  cost.   On the other hand, when the angle recovery condition is violated, then 
the convexified network may have a strictly lower optimal cost.   Moreover, this benefit can be
attained by placing phase shifters only outside an {\em arbitrary} spanning tree of the network
graph.   

There are in general many  ways to choose  phase shifter angles to convexity a network,
depending on the number and location of the phase shifters.
While placing phase shifters on each link outside a spanning tree requires
 the minimum number of phase shifters to guarantee exact relaxation, this strategy
might  require relatively large angles at some of these phase shifters.
On the other extreme, one can choose to minimize (the Euclidean norm of) the phase
shifter angles by deploying phase shifters on every link in the network.
We prove that this minimization problem is NP-hard.  Simulations suggest, however, 
that a simple heuristic  works quite well in practice.

\begin{figure}[htbp]
\centering
\includegraphics[width=0.5\textwidth]{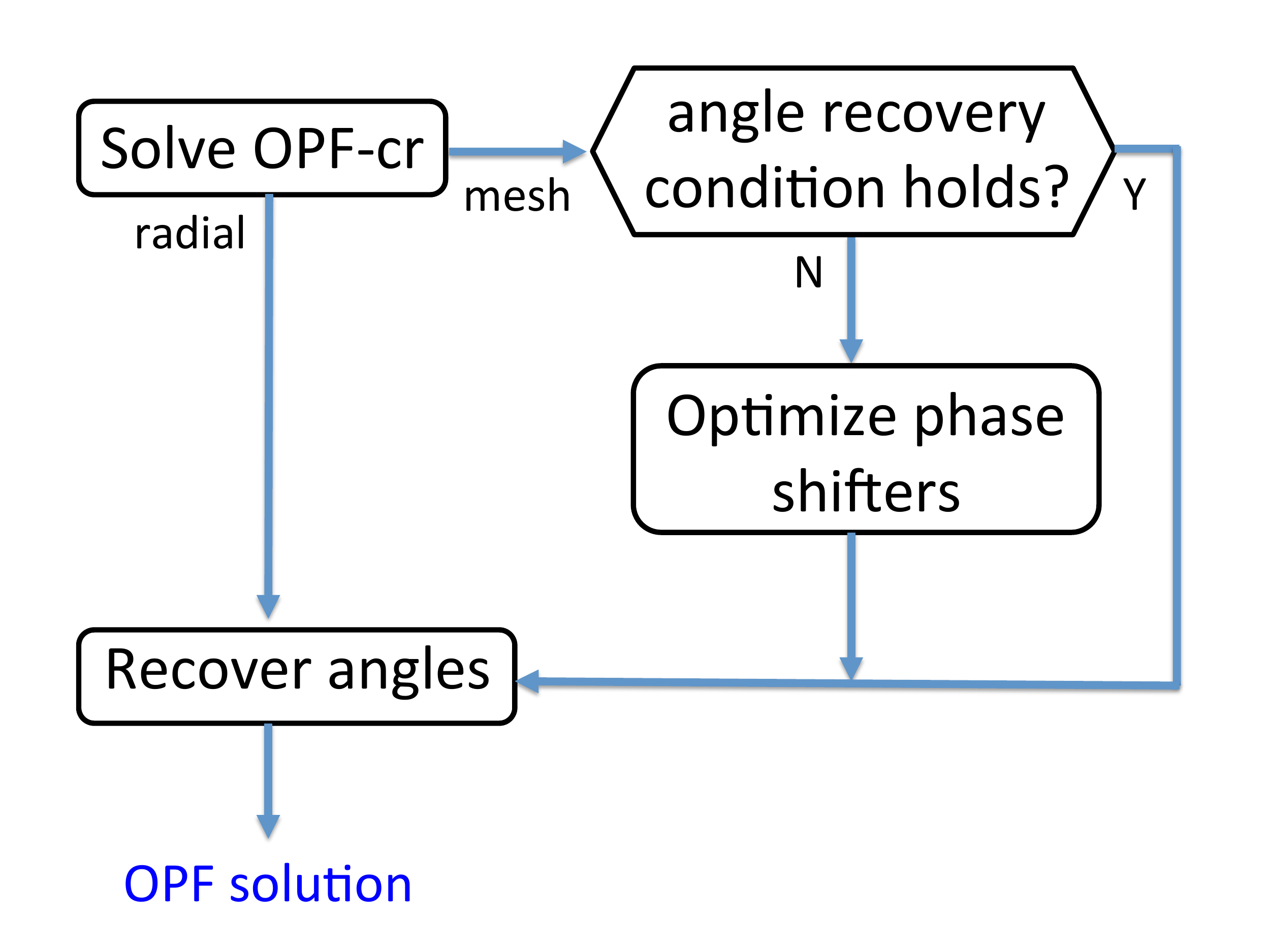}
\caption{Proposed algorithm for solving OPF with phase shifters in mesh networks.  
The details are explained in this two-part paper.}
\label{fig:strategy}
\end{figure}

These results lead to an algorithm for solving OPF when there are phase shifters in mesh
networks, as summarized in Figure \ref{fig:strategy}.
%%%%%%%%%%%%%%%%
%%%%%%%%%%%%%%%%

Since power networks in practice are very sparse, the number of lines not in a spanning tree can
be relatively small compared to the number of buses squared, as demonstrated in simulations in Section \ref{sec:sim}
 using the IEEE test systems with 14, 30, 57, 118 and 300 buses, as well as a 39-bus model of
 a New England power system and two models of a Polish power system with more than 2,000 buses.
% It is thus practical to implement the convexification proposed in this paper even for large power networks.  
Moreover, the placement of these phase shifters depends only on network topology, but not on
power flows, generations, loads, or operating constraints. 
Therefore only one-time deployment cost is required to achieve subsequent simplicity in network 
operation. 
Even when  phase shifters are not installed in the network, the optimal solution of a convex relaxation
is useful in providing a lower bound on the true optimal objective value.  This lower bound serves as a 
benchmark for other heuristic solutions of OPF.

The paper is organized as follows.   In Section \ref{sec:PS}, we extend the branch flow model
of \cite{Farivar-2013-BFM-TPS1} to include phase shifters.  In Section \ref{sec:phi}, we describe 
methods to compute phase shifter angles to map any relaxed solution to an branch flow solution.
In Section \ref{sec:opf}, we explain how to use phase shifters to simplify OPF.   In Section \ref{sec:sim},
we present our simulation results.

% \section{Convexification of mesh network}
% \label{sec:mesh}

%%%%%%%%%%%%%%%%%%%%%%%%%%%%%%%%%%%%%%%%%%%%%%%%%%%%

\section{Branch flow model with phase shifters}
\label{sec:PS}

We adopt the same notations and assumptions A1--A4 of \cite{Farivar-2013-BFM-TPS1}.

\subsection{Review: model without phase shifters}

For ease of reference, we reproduce the {\em branch flow model}
of \cite{Farivar-2013-BFM-TPS1} here:
\bq
I_{ij} & \!\!\! = \!\!\! & y_{ij} \left( V_i - V_j \right)
% \ \ \ \ \ \ \forall (i, j)\in E
\label{eq:Kirchhoff.1b}
\\
S_{ij} & \!\!\! = \!\!\! & V_i I_{ij}^*
% \ \ \ \ \ \ \forall (i, j)\in E
\label{eq:Kirchhoff.1c}
\\
 s_j & \!\!\! = \!\!\!
 	 & \!\! \sum_{k: j\rightarrow k} \!\! S_{jk}   - \!\! \sum_{i: i\rightarrow j} \!\! \left( S_{ij} - z_{ij} |I_{ij}|^2 \right) 
		+ y_j^* |V_j|^2
%  \ \ \forall j
\label{eq:Kirchhoff.1a}
\eq
Recall the set $\mathbb X(s)$  of branch flow solutions given $s$ defined in 
\cite{Farivar-2013-BFM-TPS1}:
\bq
\mathbb X(s) & \!\! \!\! := &\!\!\!\! \left\{ x := (S, I, V, s_0) \, |\, x \text{ solves 
		(\ref{eq:Kirchhoff.1b})--(\ref{eq:Kirchhoff.1a}) given $s$} \right\}
\nonumber 
\\
& & 
\label{eq:defXs}
\eq
and the set $\mathbb X$ of all branch flow solutions:
\bq
\mathbb X & := & \bigcup_{s\in \mathbb C^n}\  \mathbb X(s)
\label{eq:defX}
\eq
To simplify notation, we often use $\mathbb X$ to denote the set defined either in
\eqref{eq:defXs} or in \eqref{eq:defX}, depending on the context.
In this section we study power flow solutions and hence we fix an $s$.  All quantities, such as
$x, \hat{y}, \mathbb X, \hat{\mathbb Y}, \overline{X}, \overline{X}_T$,  are with respect to the 
given $s$, even though that is
not explicit in the notation.  In the next section, $s$ is also an optimization variable and the
sets $\mathbb X, \hat{\mathbb Y},  \overline{X}, \overline{X}_T$ are for any $s$.

Given a relaxed solution $\hat{y}$, define $\beta:= \beta(\hat{y})$  by:
\bq
\beta_{ij} & := & \angle \left( v_i - z_{ij}^* S_{ij} \right), \quad\quad (i, j)\in E
\label{eq:defb.1}
\eq
It is proved  in Theorem 2 of \cite{Farivar-2013-BFM-TPS1} that a given 
$\hat{y}$ can be mapped to a branch flow solution in $\mathbb X$ if and only if 
there exists an $(\theta, k)$ % with $\theta \in (-\pi, \pi]^n$ 
that solves
\bq
B \theta & = & \beta + 2\pi k
\label{eq:atb}
\eq
for some integer vector $k \in \mathbb N^n$.  Moreover if \eqref{eq:atb} has a solution,
then it has a countably infinite set of solutions $(\theta, k)$, but they are relatively
unique, i.e., given $k$, the solution $\theta$ is unique, and given $\theta$, the
solution $k$ is unique.
Hence \eqref{eq:atb} has a unique solution $(\theta_*, k_*)$ with $\theta_* \in (-\pi, \pi]^n$  
if and only if 
	\bq
	B_{\perp} B_T^{-1} \beta_T & = & \beta_{\perp} \quad (\text{mod } 2\pi )
	\label{eq:recoverycond}
	\eq
which is equivalent to the requirement that the (implied) voltage angle differences
sum to zero around any cycle $c$:
	\bqn
	\sum_{(i,j) \in c} \tilde{\beta}_{ij} & = & 0  \quad (\text{mod } 2\pi )
	\label{eq:cyclecond}
	\eqn
where $\tilde{\beta}_{ij} = \beta_{ij}$ if $(i,j)\in E$ and
$\tilde{\beta}_{ij} = - \beta_{ji}$ if $(j,i)\in E$.

\subsection{Model with phase shifters}

\begin{figure}[htbp]
\centering
\includegraphics[scale=0.2]{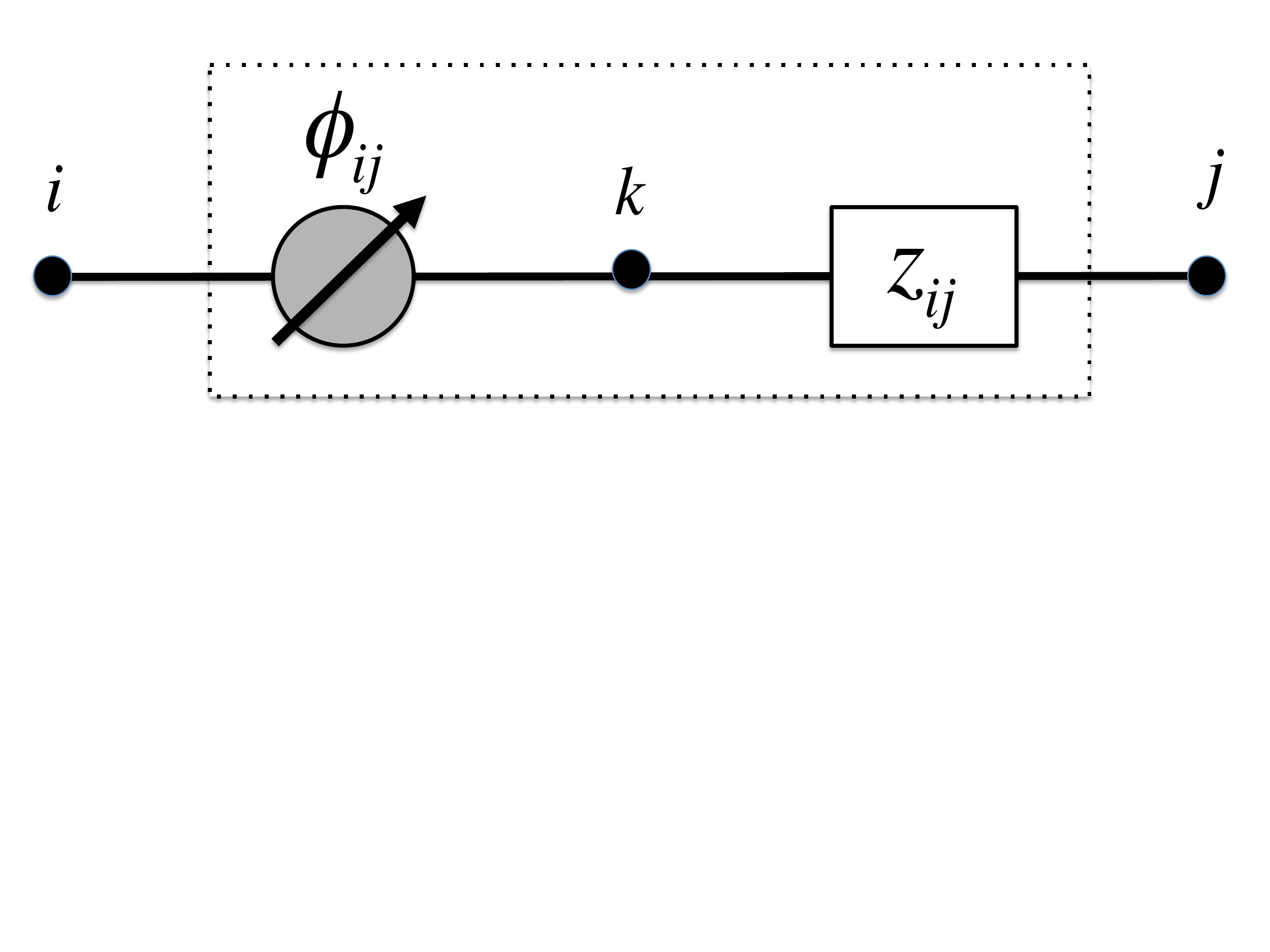}
\vspace{-0.9in}
\caption{\label{fig:PST} Model of a phase shifter in line $(i,j)$.  }
\end{figure}

Phase shifters can be traditional transformers or FACTS (Flexible AC Transmission Systems) devices.
They can increase transmission capacity and improve stability and power quality
\cite{Hingoranl-2000-FACTS-Wiley, Zhang-2006-FACTS-Springer}.
In this paper, we consider an idealized phase shifter
that  only shifts the phase angles of the sending-end voltage and current across a line,
and has no impedance nor limits on the shifted angles.
  Specifically, consider
an idealized phase shifter parametrized by $\phi_{ij}$ across line $(i, j$), as shown in
Figure \ref{fig:PST}.
As before, let $V_i$ denote the sending-end voltage.  Define $I_{ij}$ to be the {\em sending-end} current
leaving node $i$ towards node $j$.    Let $k$ be the point between the phase shifter $\phi_{ij}$
and line impedance $z_{ij}$.   Let $V_k$ and $I_k$ be the voltage at $k$ and current from $k$
to $j$ respectively.  Then the effect of the idealized phase shifter is summarized by the following modeling
assumption:
\bqn
V_k \, = \, V_i\ e^{\ii \phi_{ij}} & \text{and} &
I_{k} \, =\, I_{ij}\ e^{\ii \phi_{ij}}
\eqn
The power transferred from nodes $i$ to $j$ is still (defined to be)
$S_{ij} :=  V_i I_{ij}^*$ which, as expected, is equal to the power $V_k I_{k}^*$ from nodes $k$ to $j$
since the phase shifter is assumed to be lossless.
% Notice however that, while the sending-end current from $i$ is $I_{ij}$,
% the receiving-end current to $j$ is now $I_{ij}e^{\ii \phi_{ij}}$.
Applying Ohm's law across $z_{ij}$, we define the {\em branch flow model with phase
shifters} as the following set of equations:
\bq
I_{ij} & \!\!\! = \!\!\! & y_{ij} \left( V_i - V_j \ e^{- \ii \phi_{ij}} \right)
% \ \ \ \ \ \ \forall (i, j)\in E
\label{eq:Kirchhoff.3b}
\\
S_{ij} & \!\!\! = \!\!\! & V_i I_{ij}^*
% \ \ \ \ \ \ \forall (i, j)\in E
\label{eq:Kirchhoff.3c}
\\
 s_j & \!\!\! = \!\!\!
 	 & \!\! \sum_{k: j\rightarrow k} \!\! S_{jk}   - \!\! \sum_{i: i\rightarrow j} \!\! \left( S_{ij} - z_{ij} |I_{ij}|^2 \right) 
		+ y_j^* |V_j|^2
%  \ \ \forall j
\label{eq:Kirchhoff.3a}
\eq
Without phase shifters ($\phi_{ij}=0$), \eqref{eq:Kirchhoff.3b}--\eqref{eq:Kirchhoff.3a}
reduce to the branch flow model \eqref{eq:Kirchhoff.1b}--\eqref{eq:Kirchhoff.1a}.

The inclusion of phase
shifters modifies the network and enlargers the solution set of the (new) branch flow equations.  
Formally, let
\bq
\!\!\!\!\!\!
\overline{\mathbb X} & \! := &\!\!  \left\{ x  \, |\, x
	\text{ solves \eqref{eq:Kirchhoff.3b}--\eqref{eq:Kirchhoff.3a} for some } \phi \right\}
\label{eq:defXbar}
\eq
Unless otherwise specified, all angles should be interpreted as being modulo $2\pi$
and in  $(-\pi, \pi]$.   Hence we are primarily interested in $\phi \in (-\pi, \pi]^m$.
For any spanning tree $T$ of $G$, let ``$\phi\in T^{\perp}$'' stands for ``$\phi_{ij}=0$ for
all $(i, j)\in T$'', i.e., $\phi$ involves only phase shifters in branches not in
the spanning tree $T$.   Define
\bq
\overline{\mathbb X}_T & \!\!\! := &\!\!\! \left\{ x \, |\, x
	\text{ solves \eqref{eq:Kirchhoff.3b}--\eqref{eq:Kirchhoff.3a} for some } \phi\in T^{\perp}
									 \right\}
\nonumber
\\
\label{eq:defXbarT}
\eq
Since \eqref{eq:Kirchhoff.3b}--\eqref{eq:Kirchhoff.3a} reduce to the branch flow model
when $\phi=0$,  $\mathbb X \subseteq \overline{\mathbb X}_T \subseteq \overline{\mathbb X}$.

\section{Phase angle setting}
\label{sec:phi}

Given a relaxed solution $\hat{y}$, there are in general many  ways to
choose  angles $\phi$ on the phase shifters to recover a feasible branch flow solution 
$x \in \overline{\mathbb X}$
from $\hat{y}$.    They depend on the number and location of the phase shifters.

\subsection{Computing $\phi$}
\label{subsec:cphi}

For a network with phase shifters, we have from (\ref{eq:Kirchhoff.3b}) and \eqref{eq:Kirchhoff.3c}
\bqn
S_{ij} & = & V_i\, \frac{V_i^* - V_j^*\ e^{\ii \phi_{ij}}}{z_{ij}^* }
\eqn
leading to $V_i V_j^*\, e^{\ii \phi_{ij}}  = v_i - z_{ij}^* S_{ij}$.
Hence  $\theta_i - \theta_j =\beta_{ij} - \phi_{ij} + 2\pi k_{ij}$ for some integer $k_{ij}$.
This changes the angle recovery condition in Theorem 2 of
\cite{Farivar-2013-BFM-TPS1}  from whether there exists $(\theta, k)$ that solves
 \eqref{eq:atb} to whether there exists $(\theta, \phi, k)$ that solves
\bq
B \theta  & = & \beta - \phi + 2\pi k
\label{eq:atb.3}
\eq
for some integer vector $k\in (-2\pi, 2\pi]^m$.
The  case without phase shifters corresponds to setting $\phi = 0$.

We now describe two ways to compute $\phi$: the first minimizes the required number of phase 
shifters, and the second minimizes the size of phase angles.

\vspace{0.1in}
\subsubsection{Minimize number of phase shifters}
Our first key result implies that, given a relaxed solution 
$\hat{y} := (S, \ell, v, s_0) \in \hat{\mathbb Y}$, 
we can always recover a branch flow solution $x := (S, I, V, s_0) \in \overline{\mathbb X}$ of the convexified 
network.  Moreover it suffices to use phase shifters in branches only outside a spanning tree.   
% It extends Theorem \ref{thm.2} to the case with phase shifters.
This method requires the smallest number $(m - n)$ of phase shifters.

Given any $d$-dimensional vector $\alpha$, let $\mathcal P(\alpha)$ denote its projection onto 
$(-\pi, \pi]^d$ by taking modulo $2\pi$ componentwise.
\begin{theorem}
\label{thm.3}
Let  $T$ be {\em any} spanning tree of $G$.
Consider a relaxed solution $\hat{y} \in \hat{\mathbb Y}$
and the corresponding $\beta$ defined by (\ref{eq:defb.1}) in terms of $\hat{y}$.
\bee	
\item There exists a unique $(\theta_*, \phi_*) \in (-\pi, \pi]^{n+m}$ with $\phi_* \in T^\perp$
	such that $h_{\theta_*}(\hat{y}) \in \overline{\mathbb X}_T$, i.e.,
	 $h_{\theta_*}(\hat{y})$ is a branch flow solution of the convexified network.
	 Specifically 
	\bqn
	\theta_* & = & \mathcal{P} \left( B_T^{-1}\beta_T  \right)
	\\
	\phi_*  & = &   \mathcal{P} \left( 
			\begin{bmatrix}
			0  \\  \beta_{\perp} - B_{\perp} B_T^{-1}\beta_T
			\end{bmatrix}
			\right)
	\eqn

\item $\mathbb Y = \overline{\mathbb X} = \overline{\mathbb X}_T$ and hence
	$\hat{\mathbb Y} = \hat{h}(\overline{X}) = \hat{h}(\overline{\mathbb X}_T)$.
\eee
\end{theorem}
\begin{proof}
For the first assertion, write $\phi = [\phi_T^t\ \ \phi_{\perp}^t]^t$ and set $\phi_T = 0$.  Then
	\eqref{eq:atb.3} becomes
	\bq
	\begin{bmatrix}   B_T \\ B_{\perp}   \end{bmatrix}  \theta
	& = & 
	\begin{bmatrix}   \beta_T  \\  \beta_{\perp}   \end{bmatrix}  -  
	\begin{bmatrix}   0 \\  \phi_{\perp}   \end{bmatrix}
	+ 2\pi
	\begin{bmatrix}   k_T \\  k_{\perp}   \end{bmatrix}
	\label{eq:atb.4}
	\eq
	We now argue that there always exists a unique 
	$({\theta}_*,  {\phi}_*)$, with $\theta_* \in (-\pi, \pi]^n$, $\phi_* \in (-\pi, \pi]^m$ and 
	${\phi}_*  \in T^\perp$, that solves \eqref{eq:atb.4} for some $k\in \mathbb N^m$.
	
	The same argument as in the proof of Theorem 2 in \cite{Farivar-2013-BFM-TPS1}
	shows that a vector $({\theta}_*,  {\phi}_*, k_*)$ with $\theta_* \in (-\pi, \pi]^n$ and 
	${\phi}_*  \in T^\perp$ is a solution of \eqref{eq:atb.4} if and only if
	\bqn
		B_{\perp} B_T^{-1} \beta_T = \beta_{\perp} - [\phi_*]_{\perp} + 2\pi \left[ \hat{k}_* \right]_\perp
	\eqn
	where $\left[ \hat{k}_* \right]_\perp := [k_*]_\perp - B_{\perp} B_T^{-1} [k_*]_T$ is
	an integer vector.
	Clearly this can always be satisfied by choosing 
	\bq
	[\phi_*]_{\perp} - 2\pi \left[ \hat{k}_* \right]_\perp & = & \beta_{\perp} - B_{\perp} B_T^{-1}\beta_T
	\label{eq:phi}
	\eq
	Note that given $\theta_*$, $[k_*]_T$ is uniquely determined since $[\phi_*]_T = 0$,
	but $([\phi_*]_\perp, [k_*]_\perp)$ can be freely chosen to satisfy \eqref{eq:phi}.
	Hence we can choose the unique $[k_*]_\perp$ such that $[\phi_*]_\perp \in (-\pi, \pi]^{m-n}$.
	
	Hence we have shown that there always exists a unique 	$({\theta}_*,  {\phi_*})$, 
	with $\theta_* \in (-\pi, \pi]^n$, $\phi_* \in (-\pi, \pi]^m$ and ${\phi}_*  \in T^\perp$, 
	that solves \eqref{eq:atb.4} for some $k_* \in \mathbb N^m$.
	Moreover this unique vector $(\theta_*, \phi_*)$ is given by the formulae in the theorem.
	
The second assertion follows from assertion 1.
\end{proof}

\vspace{0.1in}
\subsubsection{Minimize phase angles}

The choice of  $(\theta_*, \phi_*)$ in Theorem \ref{thm.3} has the advantage that
it requires the minimum number of phase shifters (only on links
outside an arbitrary spanning tree $T$).   It
might however require relatively large angles $[\phi_*]_e$ at some links $e$ outside
$T$.   On the other extreme, suppose we have phase shifters on every link.
Then one can choose $(\theta_*, \phi_*)$ such that the phase shifter angles
are minimized.   

Specifically we are interested in a solution $(\theta, \phi, k)$ of 
\eqref{eq:atb.3} that minimizes $\| \mathcal P (\phi)\|^2$ where
$\|\cdot\|$  denotes the Euclidean norm of $\phi$ after taking mod $2\pi$ componentwise. 
Hence we are interested in solving the following problem: given $B, \beta$, 
\bq
\min_{\theta, \phi, k, l} & & \left\| \phi - 2\pi l \right\|^2
\label{eq:psmin.1}
\\
\text{subject to} & &  B \theta  = \beta - \phi + 2\pi k
\label{eq:psmin.2}
\eq
where $k, l \in \mathbb N^m$ are integer vectors.   
\begin{theorem}
\label{thm.4}
The problem \eqref{eq:psmin.1}--\eqref{eq:psmin.2} of minimum phase angles
is NP-hard.
\end{theorem}

\begin{proof}
Clearly the problem \eqref{eq:psmin.1}--\eqref{eq:psmin.2}  is equivalent to
the following unconstrained minimization
(eliminate $\phi$ from \eqref{eq:psmin.1}--\eqref{eq:psmin.2}):
\bq
\min_{k \in \mathbb N^m}\ \min_{\theta \in \mathbb R^n} & & \left\| ( \beta + 2\pi k) - B \theta  \right\|^2
\label{eq:psmin.3}
\eq
% where $\theta \in \mathbb R^n$ and $k\in \mathbb N^m$. 
It thus solves for a lattice point $\beta + 2\pi k$
that is closest to the range space $\{ B\theta \ |\  \theta\in \mathbb R^n\}$ of $B$,
as illustrated in Figure \ref{fig:psmin.3}.
\begin{figure}[htbp]
\centering
\includegraphics[width=0.43\textwidth]{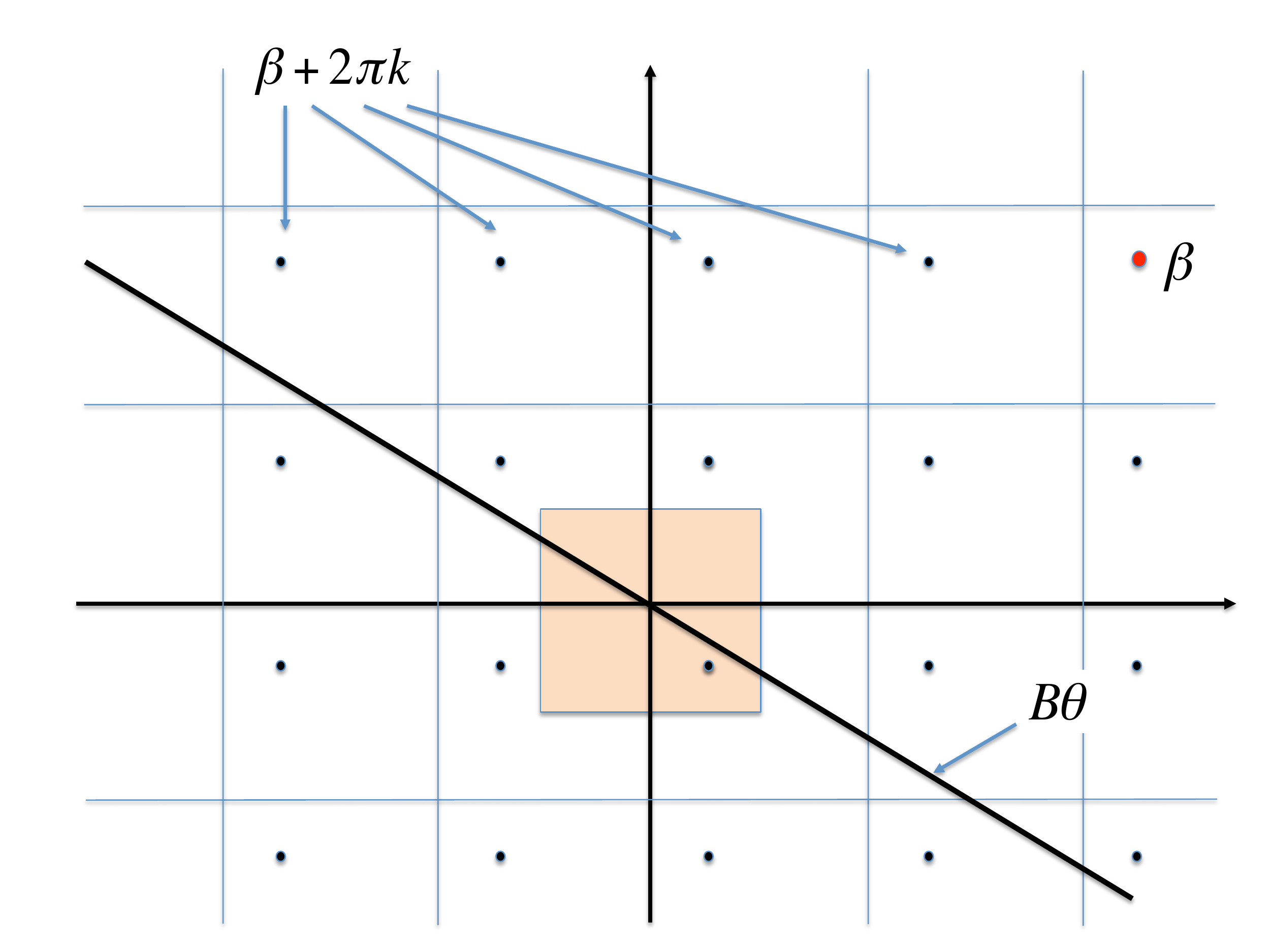}
\caption{\label{fig:psmin.3} 
Each lattice point corresponds to $2\pi k$ for an $k\in \mathbb N^m$.
The constrained optimization \eqref{eq:psmin.3} is to find a lattice point that is
closest to the range space $\{ B\theta | \theta\in \mathbb R^n\}$ of $B$. 
The shaded region around the origin is $(-\pi, \pi]^m$ and contains a
point $\beta' := \beta + 2\pi k$ for exactly one $k \in \mathbb N^m$.   Our approximate
solution corresponds to solving \eqref{eq:10} for this fixed $k$. 
}
\vspace{-0.2in}
\end{figure}

Fix any $k\in \mathbb N^m$.  Consider $\beta' := \beta + 2\pi k$ and the inner minimization
in \eqref{eq:psmin.3}:
\bq
\min_{\theta \in \mathbb R^m} & & \left\|  \beta' - B \theta  \right\|^2
\label{eq:10}
\eq
This is the standard linear least-squares estimation where $\beta'$ represents an
observed vector that is to be estimated by an vector in the range space of $B$ in order to 
minimize the normed error squared.   The optimal solution is:
\bq
\theta_*  & := & \mathcal  ( B^t B )^{-1} B^t  \beta'  
\label{eq:12}
\\
\beta' - B\theta_* & = & \left( I - B ( B^t B )^{-1} B^t  \right) \beta'
\label{eq:11}
\eq
Substituting \eqref{eq:11} and \eqref{eq:10} into \eqref{eq:psmin.3}, 
\eqref{eq:psmin.3}  becomes
\bq
\min_{k \in \mathbb N^m} & & \left\| \gamma + 2\pi A k  \right\|^2
\label{eq:psmin.4}
\eq
where $\gamma := A\beta \in \mathbb R^m$ and 
$A :=  I - B ( B^t B )^{-1} B^t $ is the orthogonal complement
of the range space of $B$.
But \eqref{eq:psmin.4} is the {\em closest lattice vector problem} and is known
to be NP-hard \cite{Ajtai1998}.\footnote{We thank Babak Hassibi for pointing
out \eqref{eq:psmin.4} is the closest lattice vector problem studied in the literature.}
\end{proof}

\begin{remark}
Since the objective function is strictly convex, the phase angles
$\phi_* = ( \beta + 2\pi k_*) - B \theta_*$ at optimality will lie in $(-\pi, \pi]^m$.
Moreover, if an optimal solution exists, then there is always an optimal solution
with $\theta_*$ in  $(-\pi, \pi]^n$: 
 if $(\theta, k)$ is  optimal  for 
\eqref{eq:psmin.3} with $\theta \not\in (-\pi, \pi]^n$, then 
by writing $k =: B\alpha + k'$ for  integer vectors $\alpha\in \mathbb N^n$, 
$k' \in\mathbb N^m$, the objective function in \eqref{eq:psmin.3} becomes
\bqn
( \beta + 2\pi k') - B (\theta - 2\pi \alpha)  
\eqn
i.e., we can always choose $(k_*, \alpha_*)$ so that $\theta_* :=  \theta - 2\pi \alpha_*$ lies
in $(-\pi, \pi]^n$ and $k = B\alpha_* + k_*$.
Therefore, given an optimal solution $(\theta, k)$ with $\theta \not\in (-\pi, \pi]^n$, 
we can find another point $(\theta_*, k_*)$ with $\theta_*= \mathcal P(\theta) \in (-\pi, \pi]^n$ 
that is also optimal.
\end{remark}

Many algorithms have been proposed to solve the closest lattice vector problem.
See \cite{Hanrot2011} for state-of-the-art algorithms.
Given $\beta$, there is a unique $k$ such that $\beta' := \beta + 2\pi k$ is in $(-\pi, \pi]^m$,
as illustrated in the shaded area of Figure \ref{fig:psmin.3}.
A simple heuristic that provides an upper bound on \eqref{eq:psmin.3} is to solve
\eqref{eq:10} for this fixed $k$.
From \eqref{eq:12}--\eqref{eq:11}, the heuristic solution is
\bqn
\theta_*  & := & \mathcal P \left( ( B^t B )^{-1} B^t  \beta'  \right)
\\
\phi_*  & := & \left( I - B ( B^t B )^{-1} B^t  \right) \beta'
\eqn
This approximate solution is illustrated in Section \ref{sec:sim} and seems to be
effective in reducing the phase shifter angles ($k=0$ in all our test cases).

\subsection{Arbitrary network of phase shifters}

More generally, consider a network with phase shifters on an arbitrary subset of
links.  Given a relaxed solution $\hat{y}$,
under what condition does there exists a $\theta$ such that the inverse projection 
$h_\theta(\hat{y})$ is a branch flow solution in $\mathbb X$?   If there is a spanning tree
$T$ such that all links outside $T$ have phase shifters, then Theorem \ref{thm.3} says 
that such a $\theta$ always exists, with an appropriate choice of phase shifter angles on
non-tree links.  Suppose no such spanning tree exists, i.e., given any spanning tree $T$,
there is a set $E_{\perp'} \subseteq E\setminus E_T$ of links that contain no phase shifters.
Let $B_{\perp'}$ and $\beta_{\perp'}$ denote the submatrix of $B$ and subvector of $\beta$,
respective, corresponding to these links.   
Then a necessary and sufficient condition for angle recovery is: there exists a spanning tree 
$T$ such that the associated $B_{\perp'}$ and $\beta_{\perp'}$ satisfy:
\bq
B_{\perp'} B_T^{-1} \beta_T  & = &  \beta_{\perp'}
\quad (\text{mod } 2\pi )
\label{eq:anps}
\eq
This condition reduces to \eqref{eq:recoverycond} if there are no phase shifters in the network
($E_{\perp'} = E\setminus E_T$) 
and is always satisfied if every link outside any spanning tree has a phase shifter
($E_{\perp'} = \emptyset$).
It requires that the  angle differences implied by $\hat{y}$ sum to zero (mod $2\pi$) around any
loop that contains no phase shifter (c.f. Theorem 2(1) and Remark 4 of \cite{Farivar-2013-BFM-TPS1}).
After such a $T$ is identified, the above two methods can be used to compute the required
phase shifts.

\subsection{Other properties}

We close this section by discussing two properties of $\phi$.   
First, the voltage angles are $\theta = \mathcal P(B_T^{-1} (\beta_T - \phi_T))$ 
and the angle recovery condition \eqref{eq:recoverycond} becomes
\bq
B_\perp B_T^{-1} ({\beta}_T  -  \phi_T)   & = & \beta_\perp - \phi_\perp
\quad (\text{mod } 2\pi)
\label{eq:recoverycond.ps}
\eq
which can always be satisfied by appropriate (nonunique) choices of $\phi$.
A similar argument to the proof of Theorem 2(2) leads to the following
interpretation of \eqref{eq:recoverycond.ps}.   For any link $(i,j)\in E$, \eqref{eq:atb.3} says
that the phase angle difference from
node $i$ to node $j$ is $\beta_{ij}$ and consists of the voltage angle difference
$\theta_i - \theta_j = \beta_{ij} - \phi_{ij}$ and the phase shifter angle $\phi_{ij}$.
Fix any link $(i, j) \in E\setminus E_T$ not in tree $T$.     
The left-hand side $\left[ B_\perp B_T^{-1} ({\beta}_T - {\phi}_T) \right]_{ij}$ of \eqref{eq:recoverycond.ps} 
represents the
sum of the voltage angle differences from node $i$ to node $j$ along the unique path in $T$,
{\em not} including the phase shifter angles along the path.  This must be equal to 
the voltage angle difference $\left[ {\beta}_\perp -{\phi}_\perp \right]_{ij}$ 
across (the non-tree) link $(i,j)$, {\em not} including the phase shifter angle across  $(i,j)$.
Therefore \eqref{eq:recoverycond.ps} has the same interpretation as before that
the voltage angle differences sum to zero (mod $2\pi$) around any cycle, though, with phase
shifters, the voltage angle differences are now $\beta_{ij} - \phi_{ij}$ instead of
$\beta_{ij}$.
This in particular leads to a relationship between any two solutions $(\theta_*, \phi_*)$
and  $(\hat{\theta}, \hat{\phi})$ of \eqref{eq:atb.3}.

In particular, let $(\theta_*, \phi_*)$ be the solution in Theorem \ref{thm.3} where $\phi_*\in T^{\perp}$,
and $(\hat{\theta}, \hat{\phi})$ any other solution.  Then applying \eqref{eq:recoverycond.ps} to
both $\phi_*$ and $\hat{\phi}$ leads to a relation between them on every basis cycle.
Specifically, let $i\rightarrow j$ be a link not in the spanning tree $T$, let $T(0\leadsto k)$ be 
the unique path in $T$ from node 0 to any node $k$.  Then for each link $i\rightarrow j$ in $E$
that is not in $T$, we have (equalities to be interpreted as mod $2\pi$)
\bqn
\left[\phi_*\right]_{ij} & = & 
\hat{\phi}_{ij}  - \sum_{(k,l)\in T(0\leadsto j)} \hat{\phi}_{kl} +  \sum_{(k,l)\in T(0\leadsto i)} \hat{\phi}_{kl} 
\\
& = & 
\beta_{ij}  - \sum_{(k,l)\in T(0\leadsto j)} \beta_{kl} +  \sum_{(k,l)\in T(0\leadsto i)} \beta_{kl} 
\eqn

%%%%%%%%%%%%%%%%%%%%%%%%%%%%%%%%%%%%%%%%%%%%%%%%%%
%%%     REMOVE THIS REMARK FOR TPS submission
%%%%%%%%%%%%%%%%%%%%%%%%%%%%%%%%%%%%%%%%%%%%%%%%%%
% \begin{remark}
% The expression for $\phi_*$ in Theorem \ref{thm.3}.1, in particular,
% \bq
% B_\perp B_T^{-1} {\beta}_T  +  \left[ \phi_* \right]_\perp   & = & \beta_\perp
% \label{eq:1}
% \eq
% is a special case of \eqref{eq:recoverycond.ps}.  It says that, for any link $(i,j)\in E\setminus E_T$ 
% not in $T$, the sum $\left[ B_\perp B_T^{-1} {\beta}_T  \right]_{ij}$ of the voltage
% angle differences (since there is no phase shifter in $T$) from node $i$ to 
% node $j$ along the unique path in $T$  plus the phase shifter angle across link $(i,j)$ must equal 
% $\left[ \beta_\perp \right]_{ij}$, the angle difference across link $(i,j)$ implied by the relaxed 
% solution $\hat{y}$.  
% If $\left[ \beta_\perp \right]_{ij} =  \left[ \theta_* \right]_i -  \left[ \theta_* \right]_j$, then 
% the sum of voltage angle differences around the (unique) basis cycle defined by $(i,j)$ is
% zero, and $\left[ \phi_* \right]_{ij} = 0$; this is the case in Theorem \ref{thm.2}.2 for link
% $(i,j)$.   Otherwise, a nonzero phase shifter angle $\left[ \phi_* \right]_{ij}$ is required to
% compensate for the angle difference.
% \end{remark}
%%%%%%%%%%%%%%%%%%%%%%%%%%%%%%%%%%%%%%%%%%%%%%%%%%
\vspace{0.1in}

Second,
Theorem \ref{thm.3} implies that given any relaxed solution $\hat{y}$, there exists a $\phi \in T^\perp$
 such that its inverse projection $x := h_{\theta}(\hat{y})$ is a branch flow solution, i.e., $(x, \phi)$ satisfies
\eqref{eq:Kirchhoff.3b}--\eqref{eq:Kirchhoff.3a}.    We now give an alternative direct construction of 
such a solution $(x, \phi)$ from any given branch flow solution $\tilde{x}$ and phase shifter setting 
$\tilde{\phi}$ that may have nonzero angles on some links in $T$.   It exhibits how the
effect of phase shifters in a tree is equivalent to changes in voltage angles.

Fix any spanning tree $T$.  Given any $(\tilde{x}, \tilde{\phi})$, partition
$\tilde{\phi}^t = [\tilde{\phi}_T \ \ \tilde{\phi}_\perp]$ with respect to $T$.  
Define $\alpha \in (-\pi, \pi]^n$ by 
$B_T\, \alpha = \tilde{\phi}_T$ or $\alpha := B_T^{-1} \, \tilde{\phi}_T$.  
Then define the mapping $(x, \phi) = g(\tilde{x}, \tilde{\phi})$ by
\bq
V_i \ := \ \tilde{V}_i e^{\ii \alpha_i},  \quad 
I_{ij} \ := \ \tilde{I}_{ij} e^{\ii \alpha_i},  \quad 
S_{ij} \ := \ \tilde{S}_{ij}
\label{eq:defg.1}
\eq
and
\bq
\phi_{ij} & := & \begin{cases}
		0  & \text{ if } (i,j) \in E_T
		\\
		\tilde{\phi}_{ij} - (\alpha_i - \alpha_j)     & \text{ if } (i,j) \in E\setminus E_T
		\end{cases}
\label{eq:defg.2}
\eq
i.e., $\phi$ is nonzero only on non-tree links.    
It can be verified that $\alpha_i - \alpha_j = \sum_{e\in T(i\leadsto j)} \tilde{\phi}_e$
where $T(i\leadsto j)$ is the unique path in tree $T$ from node $i$ to node $j$.
Note that $|V_i| = |\tilde{V}_i|$,
$|I_{ij}| = |\tilde{I}_{ij}|$ and $S = \tilde{S}$.  Hence if $h(\tilde{x})$ is a relaxed
branch flow solution, so is $h(x)$.   Moreover, the effect of phase shifters in $T$ is
equivalent to adding $\alpha_i$ to the phases of $V_i$ and $I_{ij}$.
\begin{theorem}
\label{thm:phiT}
Fix any tree $T$.  If $(\tilde{x}, \tilde{\phi})$ is a solution of
\eqref{eq:Kirchhoff.3b}--\eqref{eq:Kirchhoff.3a}, so is $(x, \phi) = g(\tilde{x}, \tilde{\phi})$
defined in \eqref{eq:defg.1}--\eqref{eq:defg.2}.
\end{theorem}
\begin{proof}
\footnote{A less direct proof is to observe that \eqref{eq:recoverycond.ps} and
$\alpha = B_T^{-1} \tilde{\phi}_T$ imply 
\bqn
B_\perp B_T^{-1} {\beta}_T & = & {\beta}_\perp - (\tilde{\phi}_\perp - B_\perp \alpha)
 \ = \ \beta_\perp - \phi_\perp
 \quad (\text{ mod} 2\pi)
\eqn
which means $(x, \phi)$ satisfies \eqref{eq:atb.3}.
}
Since $|V_i| = |\tilde{V}_i|$, $|I_{ij}| = |\tilde{I}_{ij}|$ and $S = \tilde{S}$, $(x, \phi)$ satisfies
\eqref{eq:Kirchhoff.3c}--\eqref{eq:Kirchhoff.3a}.    
For any link $(i,j)\in E_T$ in tree $T$, \eqref{eq:defg.1}--\eqref{eq:defg.2} imply
\bqn
V_i - V_j \, e^{-\ii {\phi}_{ij}} & = & 
\left( \tilde{V}_i -  \tilde{V}_j \, e^{- \ii (\alpha_i - \alpha_j) } \right) e^{\ii \alpha_i} 
\\
& = & \left( \tilde{V}_i -  \tilde{V}_j \, e^{- \ii \tilde{\phi}_{ij} } \right) e^{\ii \alpha_i} 
\eqn
where the second equality follows from $B_T \, \alpha = \tilde{\phi}_T$.  
For any link $(i,j)\in E\setminus E_T$ not in $T$, \eqref{eq:defg.1}--\eqref{eq:defg.2} imply
\bqn
V_i - V_j \, e^{-\ii {\phi}_{ij}} & = & 
\left( \tilde{V}_i -  \tilde{V}_j \, e^{- \ii \tilde{\phi}_{ij} } \right) e^{\ii \alpha_i} 
\eqn
But $\left( \tilde{V}_i -  \tilde{V}_j \, e^{- \ii \tilde{\phi}_{ij} } \right) = \tilde{I}_{ij}$ since 
$(\tilde{x}, \tilde{\phi})$ satisfies \eqref{eq:Kirchhoff.3b}.
Therefore  $V_i - V_j \, e^{-\ii {\phi}_{ij}} = I_{ij}$, 
i.e., $(x, \phi)$ satisfies \eqref{eq:Kirchhoff.3b}  on every link.
\end{proof}

%%%%%%%%%%%%%%%%%%%%%%%%%%%%%%%%%%%%%%%%%%%%%%%%%%
%%%     REMOVE THIS REMARK FOR TPS submission
%%%%%%%%%%%%%%%%%%%%%%%%%%%%%%%%%%%%%%%%%%%%%%%%%%%
%
% \begin{remark}
% A less direct proof of Theorem \ref{thm:phiT} is as follows.
% By \eqref{eq:recoverycond.ps}, we must have 
% \bqn
% B_\perp B_T^{-1} ({\beta}_T - \tilde{\phi}_T) & = & {\beta}_\perp - \tilde{\phi}_\perp
% \eqn
% where ${\beta} := {\beta}(h(\tilde{x})) = \beta(h(x))$ is defined by \eqref{eq:defb.1} and
% the relaxed solutions $h(\tilde{x})$ and $h(x)$ are identical.
% Substituting $\alpha = B_T^{-1} \tilde{\phi}_T$ and rearranging, this becomes
% (using \eqref{eq:defg.2})
% \bqn
% B_\perp B_T^{-1} {\beta}_T & = & {\beta}_\perp - (\tilde{\phi}_\perp - B_\perp \alpha)
%  \ = \ \beta_\perp - \phi_\perp
% \eqn
% This is simply the necessary and sufficient condition for $(x, \phi)$ to satisfy \eqref{eq:atb.3}.
% \end{remark}
%%%%%%%%%%%%%%%%%%%%%%%%%%%%%%%%%%%%%%%%%%%%%%%%%%

%%%%%%%%%%%%%%%%%%%%%%%%%%%%%%%%%%%%%%%%%%%%%%%%%%%%
\section{OPF in convexified network}
\label{sec:opf}

Theorem \ref{thm.3} suggests using phase shifters to convexify a mesh network so that 
 {\em any} solution of OPF-ar can be mapped into an optimal solution of OPF of the convexified
 network.
% By assumption, the original branch flow equations have a solution, i.e., $\mathbb X$ is nonempty,
% and therefore an optimal solution $(x_*, s_*)$ of OPF always exists.
% When a solution $(\hat{y}, s)$ of OPF-ar produces a relaxed branch flow solution $\hat{y}$ that
%  lies outside $\hat{h}(\mathbb X)$, $(\hat{y}, s)$  is clearly spurious as no optimal solution of OPF 
% can be recovered from it.   Why then should we modify the network to fit this spurious solution rather
% than trying to solve the original OPF?  One justification is that OPF is often difficult
% to solve.  
Convexification thus modifies a NP-hard problem into a simple problem without loss
in optimality; moreover this requires an one-time deployment cost for subsequent 
operational simplicity, as we now show.

We will compare four OPF problems: the original OPF 
defined in \cite{Farivar-2013-BFM-TPS1}:
\\
\noindent
\textbf{OPF}:
\bqn
\min_{x, s} & & f \left( \hat{h}(x), s \right)
\label{eq:opf1.a}
\\
\text{subject to} & &  x \in \mathbb X, \quad (S, v, s_0, s) \in \mathbb S
\label{eq:opf1.c}
\eqn
the relaxed OPF-ar defined in \cite{Farivar-2013-BFM-TPS1}:
\\
\noindent
\textbf{OPF-ar}:
\bqn
\min_{x, s} & & f \left( \hat{h}(x), s \right)
\label{eq:opf2.a'}
\\
\text{subject to} & &  x \in \mathbb Y, \quad (S, v, s_0, s) \in \mathbb S
\label{eq:opf2.c'}
\eqn
the following problem where there is a  phase shifter on every line $(\phi\in (-\pi, \pi]^m$):
\\
\noindent
\textbf{OPF-ps:}
\bqn
\min_{x, s, \phi} & & f \left( \hat{h}(x), s \right)
\label{eq:opf4a.a}
\\
\text{subject to} & &  x \in \overline{\mathbb X}, \ (S, v, s_0, s) \in \mathbb S
\label{eq:opf4a.b}
\eqn
and the problem where, given any spanning tree $T$, there are phase shifters only 
outside $T$:
\\
\noindent
\textbf{OPF-ps(T):}
\bqn
\min_{x, s, \phi} & & f \left( \hat{h}(x), s \right)
\label{eq:opf5a.a}
\\
\text{subject to} & &  x \in \overline{\mathbb X}_T, \  (S, v, s_0, s) \in \mathbb S, \ \phi \in T^\perp
\label{eq:opf5a.b}
\eqn
Let the optimal values of OPF, OPF-ar, OPF-ps, and OPF-ps(T) be $f_*$, $f_{ar}$, $f_{ps}$,
and $f_{T}$ respectively.

Theorem \ref{thm.3} implies that
$\mathbb X \subseteq  \mathbb Y = \overline{\mathbb X} = \overline{\mathbb X}_T$
for any spanning tree $T$.  Hence we have
\begin{corollary}
\label{cor.1}
For any spanning tree $T$, 
 $f_* \geq f_{ar} = f_{ps} = f_{T}$, with equality if  there is a solution $(\hat{y}_{ar}, s_{ar})$ of 
 OPF-ar that satisfies \eqref{eq:recoverycond}.
\end{corollary}

\vspace{0.1in}
Corollary \ref{cor.1} has several important implications:
\bee
\item Theorem 1 in  \cite{Farivar-2013-BFM-TPS1}
	 implies that we can solve OPF-ar efficiently through conic relaxation to obtain
	a relaxed solution $(\hat{y}_{ar}, s_{ar})$.
	An optimal solution  of OPF may or may not be recoverable from it.
	If $\hat{y}_{ar}$ satisfies the angle recovery condition
	 (\ref{eq:recoverycond}) with respect to $s_{ar}$, then Theorem 2 in  \cite{Farivar-2013-BFM-TPS1} 
	guarantees a unique $\theta_* \in (-\pi, \pi]^n$ such that the inverse projection 
	$(\hat{h}_{\theta_*}(\hat{y}_{ar}), s_{ar})$ is indeed optimal for OPF.

\item	In this case, Corollary \ref{cor.1} implies that adding any phase shifters to the network cannot further reduce
	the cost  since $f_* = f_{ar} = f_{ps}$.
	
\item If (\ref{eq:recoverycond}) is not satisfied, then $\hat{y}_{ar} \not\in \hat{h}(\mathbb X)$ and there is no 
	inverse projection that can recover an optimal solution of OPF from $(\hat{y}_{ar}, s_{ar})$.
	In this case, $f_* \geq f_{ar}$.  
	Theorem \ref{thm.3} implies that if we allow phase shifters, we can always attain
	$f_{ar}=f_{ps}$ 	with the relaxed solution $(\hat{y}_{ar}, s_{ar})$, with potentially strict improvement 
	over the network without phase shifters (when $f_* > f_{ar}$).

\item Moreover, Corollary \ref{cor.1} implies that such benefit can be achieved with phase shifters only
	in branches outside an arbitrary spanning tree $T$.
\eee

\vspace{0.1in}
\begin{remark}
The choice of the spanning tree $T$ does not affect the conclusion of the theorem.
Different choices of $T$ correspond to different choices of $n$ linearly independent
rows of  $B$ and the resulting decomposition of $B$ and $\beta$
into $B_T$ and $\beta_T$.   Therefore $T$ determines the phase angles $\theta_*$ and
$\phi_*$ according to the formulae in the theorem.
Since the objective $f(\hat{h}(x), s)$ of OPF is independent of the phase angles
$\theta_*$, for the same relaxed solution $\hat{y}$, OPF-ps achieves the same 
objective value regardless of the choice of $T$.
\end{remark}

\section{Simulations}
\label{sec:sim}

For radial networks, results in Part I (Theorem 4) guarantees that both the angle relaxation and the conic
	relaxation are exact. 
For mesh networks, the angle relaxation may be inexact
 and phase shifters may be needed to implement a solution of the
conic relaxation.
We now explore through numerical experiments the following questions:
\bi
\item  How many phase shifters are typically needed to convexify a mesh network?
\item  What are typical phase shifter angles to implement an optimal solution for the convexified network?
\ei

\textbf{Test cases.}
We explore these questions using the IEEE benchmark systems with 14, 30, 57, 118 and 300 buses, as well
as a 39-bus model of a New England power system and two models of a Polish power system
with 2,383 and 2,737 buses.
The data for all the test cases  were extracted from the library of built in models
of the MATPOWER toolbox  \cite{Zimmerman09} in Matlab.
The test cases involve constraints on bus voltages  as well as limits on the real and reactive power
generation at every generator bus.   The New England and the Polish power systems also involve
MVA limits on branch power flows.
All these systems are mesh networks, but very sparse.

\textbf{Objectives.}
We solve the test cases for two scenarios:
\bi
\item {\em Loss minimization.}
In this scenario, the objective is to minimize the total active power loss of the circuit given
 constant load values, which is equivalent to minimizing the total active power generation.
The results are shown in Table \ref{simulationsofloss}.

\item {\em Loadability maximization.}
In this scenario, the objective  is to determine the maximum possible load increase in the system while
satisfying the generation, voltage and line constraints. We have assumed all real and reactive loads
grow uniformly, i.e., by a constant multiplicative factor called the {\em max loadability}
 in Table \ref{simulationsofloadability}.
\ei

\begin{table*}[t]
\footnotesize
\centering
\begin{tabular}{cr|r@{.}l|r@{.}l|r|r|c|c|}
  \cline{3-10}
 & & \multicolumn{2}{c}{No PS} &  \multicolumn{6}{|c|} {With phase shifters (PS)} \\
 \hline
\multicolumn{1}{|c|}{Test cases}  & \multicolumn{1}{c|}{\# links}
	&  \multicolumn{2}{c}{Min loss}  &  \multicolumn{2}{|c|}{Min loss}
	& \multicolumn{1}{c|}{\# required PS}  & \multicolumn{1}{c|}{\# active PS}
	& \multicolumn{1}{c|}{Min \#PS ($^\circ$)} & \multicolumn{1}{c|}{Min $\|\phi\|^2$ ($^\circ$)} \\
	
\multicolumn{1}{|c|}{ }    & ($m$) \ &  \multicolumn{2}{c}{(OPF, MW)} & \multicolumn{2}{|c|}{(OPF-cr, MW)}
	&  \multicolumn{1}{c|}{$(m-n)$} & $|\phi_i|>0.1^\circ$ & $[\phi_{\min},\phi_{\max}]$  & $[\phi_{\min},\phi_{\max}]$\\
	
  \hline
  \hline
\multicolumn{1}{|c|}{IEEE 14-Bus}   & 20  & \ \ \ 0 & 546  & \ \ \ 0&545 & 7  \ \ \ (35\%) & 2   \ \ \ (10\%) &  $[-2.09, \ 0.58]$ &  [ -0.63, 0.12]\\ %& $[-0.70, \ 0.02]$ \\
\multicolumn{1}{|c|}{IEEE 30-Bus}  & 41  & \ \ \ 1 & 372 & \ \ \ 1&239 & 12  \ \ \ (29\%) & 3   \  \, \ \ (7\%) &  $[-0.20, \ 4.47]$  &  [-0.95, 0.65]\\ %& $[-0.03, \ 0.50]$\\
\multicolumn{1}{|c|}{IEEE 57-Bus}   & 80  &\ \  \ 11 & 302  & \ \ \ 10&910 & 24  \ \ \ (30\%) & 19  \ \ \ (24\%) & $[-3.47, \ 3.15]$ &  [-0.99, 0.99]\\ %&  $[-0.58, \ 0.38]$ \\
\multicolumn{1}{|c|}{IEEE 118-Bus}  & 186   & \ \ \ 9 & 232  &\ \ \  8&728 & 69 \ \ \ (37\%)  & 36  \ \ \ (19\%) & $[-1.95, \ 2.03]$  & [-0.81, 0.31]\\ %& $[-0.39, \ 0.28]$ \\
\multicolumn{1}{|c|}{IEEE 300-Bus}  & 411 & \ \ \ 211&871  & \ \ \ 197&387 & 112 \ \ \ (27\%) & 101 \ \ \ (25\%) &  $[-13.3, \ 9.40]$ &  [-3.96, 2.85 ]\\ %&   $[-3.96, \ 1.68]$\\
\multicolumn{1}{|c|}{New England 39-Bus}  &  46& \ \ \ 29&915& \ \ \ 28&901  & 8  \ \ \ (17\%) & 7   \ \ \ (15\%) &  $[-0.26, \ 1.83]$  &  [-0.33, 0.33]\\ %& $[-0.04, \ 0.26]$ \\
\multicolumn{1}{|c|}{Polish (case2383wp)}  & 2,896 & \ \ \ 433&019 &  \ \ \ 385&894 & 514  \ \ \ (18\%)  &  373  \ \ \ (13\%)  &  $[-19.9, \ 16.8]$  & [-3.07, 3.23]\\ %& $[-1.68, \ 0.93]$\\
\multicolumn{1}{|c|}{Polish (case2737sop)}  & 3,506 &  \ \ \ 130&145 & \ \ \ 109&905  &  770  \ \ \ (22\%)  &  395 \ \ \ (11\% ) &  $[-10.9, \ 11.9]$ & [ -1.23, 2.36]\\ %&  $[-0.93, \ 0.71]$\\
  \hline
\end{tabular}
  \caption{Loss minimization.  Min loss without phase shifters (PS) was computed using SDP relaxation of OPF;
  % (\ref{eq:opf1.a})--(\ref{eq:opf1.c}); 
  min loss with phase shifters was computed using SOCP relaxations OPF-cr
  % (\ref{eq:opf3.a})--(\ref{eq:opf3.d}) 
  of OPF-ar.
  The ``(\%)'' indicates the number of PS as a percentage of \#links.}
  \label{simulationsofloss}
  \end{table*}

\begin{table*}[t]
\footnotesize
\centering
\begin{tabular}{c|c|c|r|r|c|c|c|}
  \cline{2-8}
 & \multicolumn{1}{c}{No PS} &  \multicolumn{6}{|c|} {With phase shifters (PS)} \\
 \hline
\multicolumn{1}{|c|}{Test cases}  &   \multicolumn{1}{c}{Max load.}
	 &  \multicolumn{1}{|c|}{Max load.} & \multicolumn{1}{c|}{\# required PS}  & \multicolumn{1}{c|}{\# active PS}
	 & \multicolumn{1}{c|}{Min \#PS ($^\circ$) } & \multicolumn{1}{c|}{Min $\|\phi\|^2$ ($^\circ$)} 
	 & \multicolumn{1}{c|}{Simu. time}  \\
\multicolumn{1}{|c|}{ }     &  \multicolumn{1}{c}{(OPF)} & \multicolumn{1}{|c|}{(OPF-cr)}
	&  \multicolumn{1}{c|}{$(m-n)$} & $|\phi_i|>0.1^\circ$ & $[\phi_{\min},\phi_{\max}]$  
	& $[\phi_{\min},\phi_{\max}]$	& (seconds) \\
  \hline
  \hline

\multicolumn{1}{|c|}{IEEE 14-Bus}  & 195.0\%  & 195.2\%  & 7  \ \ \ (35\%)  & 6 \ \ \ (30\%)   &  $[-0.51, \ 1.35]$  
	& [-0.28, \ 0.23]	&  $1.92$\\
\multicolumn{1}{|c|}{IEEE 30-Bus}    & 156.7\%  &  158.7\%  & 12  \ \ \ (29\%) & 9  \ \ \ (22\%) &  $[-0.42, \ 12.4]$ 
	& [-2.68, \ 1.86]	&  $3.86$ \\
\multicolumn{1}{|c|}{IEEE 57-Bus}   & 108.2\%  & 118.3\% & 24  \ \ \ (30\%) & 24  \ \ \ (30\%) &  $[-13.1, \ 23.2]$ 
	& [-4.12, \ 4.12]	&  $7.13$ \\
\multicolumn{1}{|c|}{IEEE 118-Bus}  & 203.7\%  & 204.9\% &  69 \ \ \ (37\%)   & 64  \ \ \ (34\%) &  $[-8.47, \ 17.6]$ 
	& [-4.61, \ 5.36]	&  $15.96$\\
\multicolumn{1}{|c|}{IEEE 300-Bus} & 106.8\%  & 112.8\% & 112 \ \ \ (27\%) & 103 \ \ \ (25\%) &  $[-15.0, \ 16.5]$ 
	& [-4.28, \ 6.31]	&  $34.6$\\
\multicolumn{1}{|c|}{New England 39-Bus}    &  109.1\%  &  117.0\%   & 8  \ \ \ (17\%)  &  5  \ \ \ (11\%) & $[-1.02, \ 1.28]$ 
	& [-0.28, \ 0.18]	&  $2.82$\\
\multicolumn{1}{|c|}{Polish (case2383wp)} & 101.4\% &   106.6\% & 514  \ \ \ (18\%)  &  435 \ \ \ (15\%)  &  $[-19.6, \ 19.4]$ 
	& [-4.06, \ 4.32]	&  $434.5$\\
\multicolumn{1}{|c|}{Polish (case2737sop)}  &  127.6\% &    132.5\%   &  770  \ \ \ (22\%)  &  420  \ \ \ (12\%)
			&  $[-13.9, \ 17.1]$  &	[-2.07, \ 3.62]		&  $483.7$ \\
  \hline
\end{tabular}
  \caption{Loadability maximization.  Max loadability without phase shifters (PS) was computed using
  SDP relaxation of OPF; 
  % (\ref{eq:opf1.a})--(\ref{eq:opf1.c}); 
  max loadability with phase shifters was computed using SOCP relaxations OPF-cr
 % (\ref{eq:opf3.a})--(\ref{eq:opf3.d}) 
 of OPF-ar.
  The ``(\%)'' indicates the number of PS as a percentage of \#links.}
 \label{simulationsofloadability}
  \end{table*}

\textbf{Solution methods.}
We use the ``SEDUMI" solver in Matlab \cite{sedumi}.
We first solved the SOCP relaxation OPF-cr  % (\ref{eq:opf3.a})--(\ref{eq:opf3.c}) 
for a solution $(\hat{y}, s)$ of OPF-ar.
In all test cases, equality was attained at optimality for the second-order cone constraint, and hence
OPF-cr was exact, as Theorem 1 in  \cite{Farivar-2013-BFM-TPS1} would suggest.
Recall however that the load values were constants in all the test cases.  Even though this
 violated our condition that there are no upper bounds on the loads
% (c.f. (7) of \cite{Farivar-2012-BFM-TPS1}), 
% (c.f. \eqref{oversatisfaction}), 
OPF-cr turned out to be exact with respect to OPF-ar in all cases.
This confirms that the no-upper-bound condition is sufficient but not necessary for the conic
relaxation to be exact.

Using the solution $(\hat{y}, s)$ of OPF-ar,
we checked if the angle recovery condition \eqref{eq:recoverycond} was satisfied.   
% If it were then we could recover the voltage angles $\theta$  and an optimal solution
% $(h_{\theta}(\hat{y}), s)$ of the original OPF.
In all  test cases, the angle recovery condition failed and hence no $(h_{\theta}(\hat{y}), s)$ was
feasible for OPF without phase shifters.
We computed the phase shifter angles $\phi$ using both methods explained in 
Section \ref{subsec:cphi}
and the corresponding unique $(h_{\theta}(\hat{y}), s)$ that was an optimal
solution of OPF for the {convexified} network.
For the first method that minimizes the number of required phase shifters, 
% to place the phase shifters, 
we have used a minimum spanning tree of the network where the weights
on the lines are their reactance values.    
For the second method, we solve an approximation to the angle minimization that optimizes
over $\theta$ for the fixed $k$ that shifts $\beta$ to $(-\pi, \pi]^m$.

In Tables \ref{simulationsofloss} and \ref{simulationsofloadability},
we report the number $m-n$ of phase shifters potentially required, the number of active phase shifters 
(i.e., those with a phase angles greater than  $0.1^\circ$), and the range of the phase angles at optimality
using both methods.
In Table \ref{simulationsofloadability}, we also report the simulation time on an Intel 1.8 GHz Core i5 CPU.

We  report the optimal objective values of OPF with and without phase shifters
in Tables \ref{simulationsofloss} and \ref{simulationsofloadability}.
The optimal values of OPF without phase shifters were obtained by implementing the SDP formulation
and relaxation proposed in \cite{Bose2011} for solving OPF.
In all test cases, the solution matrix was of rank one and hence the SDP relaxation was exact.  
Therefore the values reported here are indeed optimal for OPF.
% (\ref{eq:opf1.a})--(\ref{eq:opf1.c}).

%%%%%%%%%%%%%%%%%%%%%%%%%%%%%%%%%%%%%%%%%%%%%%
%%%%
%%%%
%%%%   The following paragraph is removed to fit in 8 pages.
%%%%   Should include after first revision.
%%%%
%%%%
%%%%%%%%%%%%%%%%%%%%%%%%%%%%%%%%%%%%%%%%%%%%%%
The SDP relaxation requires the addition of small resistances ($10^{-6}$ pu) to every link that has a zero
resistance in the original model, as suggested in \cite{Lavaei2012}.
This addition is, on the other hand, not required for the SOCP relaxation: OPF-cr is tight with respect to
OPF-ar with or without this addition.
For comparison, we report the results where the same resistances are added for both
the SDP and SOCP relaxations.

\textbf{Summary.}   From Tables \ref{simulationsofloss} and \ref{simulationsofloadability}:
\bee
\item Across all test cases, the convexified networks
	have higher performance (lower minimum loss and higher maximum loadability) than the original
	networks.   More important than the modest performance improvement is design
	for simplicity: it guarantees an efficient solution for OPF.

\item The networks are (mesh but) very sparse, with the ratios $m/(n+1)$ of the
	number of lines to the number of buses  varying from 1.2 to 1.6 (Table \ref{simulationsofloss}).
	The numbers $m-n$ of phase shifters {potentially} required on every link outside a spanning
	tree for convexification vary from 17\% of the numbers $m$ of links to 37\%.

\item The numbers of active phase shifters in the test cases vary from 7\% of the numbers $m$ of
	links to 25\% for loss minimization, and 11\% to 34\% for loadability maximization.
	The phase angles required at optimality is no more than $20^\circ$ in magnitude with
	the minimum number of phase shifters.    With the maximum number of phase shifters,
	the range of the phase angles is much smaller (less than $7^\circ$).
	
\item The simulation times range from a few secs to mins. This is much faster than SDP relaxation. 
	Furthermore they appear linear in network size.
\eee

\normalsize

\section{Conclusion}
\label{sec:c}

We have presented a branch flow model and demonstrated how it can be used for the analysis and optimization
of mesh as well as radial networks.    Our results confirm that radial networks are computationally
much simpler than mesh networks.
For mesh networks, we have proposed a simple way to convexify them using phase shifters 
that will render them computationally as simple as radial networks for power flow solution and
optimization.   The addition of phase shifters thus convert a nonconvex problem into
a different, simpler problem.

We have proposed a solution strategy for OPF that consists of two steps:
\bee
\item Compute a relaxed solution of OPF-ar by solving its conic relaxation OPF-cr.
\item Recover from a relaxed solution an optimal solution of the original OPF using an angle 
	recovery algorithm.
\eee
We have proved that, for radial networks, both steps are always exact, provided there are
no upper bounds on loads,  so this strategy guarantees
a globally optimal solution.   For mesh networks the angle recovery condition may not hold
but can be used to check if a given relaxed solution is globally optimal.
%
% which is easily checkable, can fail.   However, with judiciously placed phase shifters, both
% relaxations will always be exact for the convexified network and the optimal cost of the
% convexified network will be no higher and potentially strictly lower than the original network.
% Moreover phase shifters are needed only on lines outside an arbitrary spanning tree of the
% network graph.  

Since practical power networks are very sparse, the number of required phase shifters
may be relatively small. 
% It is thus practical to implement the convexification proposed in this paper even for large networks. 
Moreover, their placement  depends only on 
network topology, but not on power flows, generations, loads, or operating constraints. 
Therefore an one-time deployment cost is required to achieve the subsequent simplicity in 
network and market operations.
% Indeed convexity is important not just for operation of the engineering infrastructure, but also
% prices, market operation, investment and policy decisions.	

\vspace{0.2in}
\section*{Acknowledgment}
We are grateful to S.  Bose, K. M. Chandy, L. Gan, and B. Hassibi of Caltech,
and C.  Clarke, M. Montoya, and R.  Sherick
of the Southern California Edison.   We acknowledge the support of
NSF through NetSE grant CNS 0911041, DoE's
ARPA-E through grant DE-AR0000226,  the National Science Council
of Taiwan (R. O. C.) through grant NSC 101-3113-P-008-001,
Southern California Edison, 
the Resnick Institute of Caltech, 
Cisco, and the Okawa Foundation.
% Jeff Gooding, Russel Neal, Percy Haralson, Michael Montoya, Bryan Pham, Robert Yinger and Marshall Parsons in the Advanced Technology Division of
%the Southern California Edison's (SCE) Transmission and Distribution Business Unit (TDBU) for multiple fruitful discussions.
%

\vspace{0.2in}
% \bibliographystyle{unsrt}
% \bibliography{../../../PowerRef-201202.bib}

\vspace{0.2in}
\begin{IEEEbiography}[{\includegraphics[width=1in,height=1.25in,clip,keepaspectratio]{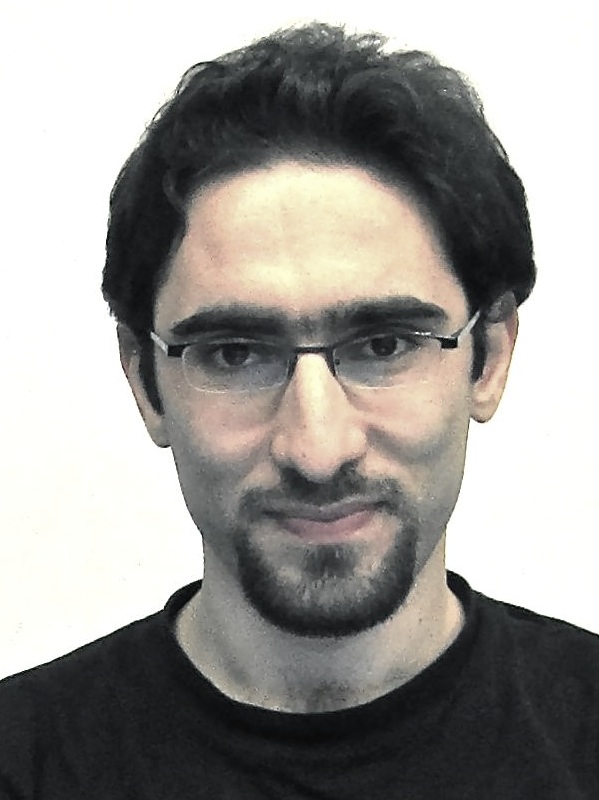}}]{Masoud Farivar}
 is a Ph.D. student of electrical
engineering at California Institute of Technology,
where he is a fellow of the Resnick Sustainability Institute.
Since June of 2010, he has also been working in the Advanced
Technology division of Southern California Edison on
a number of smart grid projects. Masoud received his
B.S. degree with a double major in electrical engineering
and computer science from Sharif University of Technology,
Tehran, Iran, in 2009.
\end{IEEEbiography}

\begin{IEEEbiography}[{\includegraphics[width=1in,height=1.25in,clip,keepaspectratio]{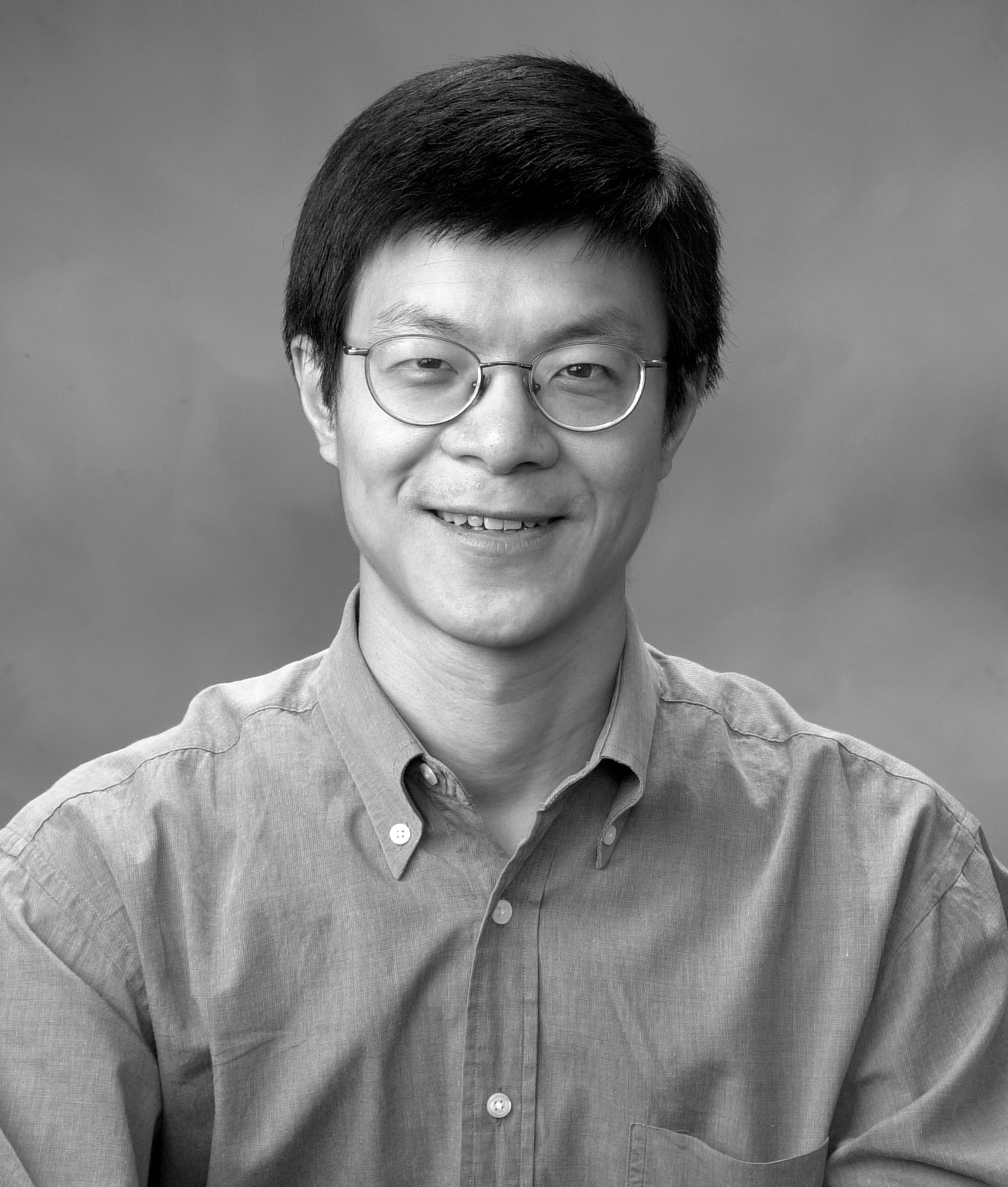}}]{Steven H. Low (F'08)}
is a Professor of the Computing \& Mathematical
Sciences and Electrical
Engineering Departments at Caltech,
and an adjunct professor of both
Swinburne University, Australia and
Shanghai Jiao Tong University, China.
Before that, he was with AT\&T Bell Laboratories, Murray
Hill, NJ, and the University of Melbourne, Australia.
He was a co-recipient of IEEE best paper awards,
the R\&D 100 Award, and an Okawa
Foundation Research Grant.  He was a
member of the Networking and Information Technology
Technical Advisory Group for the US President's
Council of Advisors on Science and Technology (PCAST)
in 2006.
He is a Senior Editor of the
IEEE Journal on Selected Areas in Communications 
(and the  mentor for the annual JSAC series on Smart Grid Communications),
a Senior Editor of the IEEE Trans. Control of Network Systems,
and a Steering Committee Member of the IEEE Trans. Network Science
\& Engineering. 
He also served on the editorial boards
of IEEE/ACM Transactions on Networking, IEEE Transactions
on Automatic Control, ACM Computing Surveys,
Computer Networks Journal, and 
NOW Foundations and Trends in Networking.  
He received his B.S. from Cornell and PhD from Berkeley,
both in EE.
\end{IEEEbiography}